\documentclass[sn-basic]{sn-jnl}

\usepackage{url}

\usepackage{hyperref}
\usepackage{subcaption}
\usepackage{array}
\usepackage{siunitx}
\usepackage{multicol}
\usepackage{times,amsmath}
\usepackage{latexsym,amssymb,lscape,multirow}
\usepackage{bm}

\graphicspath{{figures/}}

\jyear{2021}%

\theoremstyle{thmstyleone}%
\theoremstyle{thmstyletwo}%

\theoremstyle{thmstylethree}%

\begin{document}

\title[Tracking Vaccination Discussions on Mumsnet]{Tracking the Structure and Sentiment of Vaccination Discussions on Mumsnet}


\author*[1,3]{\fnm{Miguel} \sur{E. P. Silva}}\email{miguel.p.silva@inesctec.pt}

\author[1]{\fnm{Rigina} \sur{Skeva}}

\author[2]{\fnm{Thomas} \sur{House}}

\author[1]{\fnm{Caroline} \sur{Jay}}

\affil[1]{\orgdiv{Department of Computer Science}, \orgname{University of Manchester}, \orgaddress{\street{Oxford Road}, \city{Manchester}, \postcode{M13 9PY}, \state{England}, \country{United Kingdom}}}

\affil[2]{\orgdiv{Department of Mathematics}, \orgname{University of Manchester}, \orgaddress{\street{Oxford Road}, \city{Manchester}, \postcode{M13 9PY}, \state{England}, \country{United Kingdom}}}

\affil[3]{\orgdiv{LIAAD}, \orgname{INESC-TEC}, \orgaddress{\city{Porto}, \country{Portugal}}}


\abstract{Vaccination is one of the most impactful healthcare interventions in terms of lives saved at a given cost, leading the anti-vaccination movement to be identified as one of the top 10 threats to global health in 2019 by the World Health Organization. This issue increased in importance during the COVID-19 pandemic where, despite good overall adherence to vaccination, specific communities still showed high rates of refusal. Online social media has been identified as a breeding ground for anti-vaccination discussions.
  In this work, we study how vaccination discussions are conducted in the discussion forum of Mumsnet, a United Kingdom based website aimed at parents. 
By representing vaccination discussions as networks of social interactions, we can apply techniques from network analysis to characterize these discussions, namely network comparison, a task aimed at quantifying similarities and differences between networks. Using network comparison based on graphlets -- small connected network subgraphs -- we show how the topological structure vaccination discussions on Mumsnet differs over time, in particular before and after COVID-19. We also perform sentiment analysis on the content of the discussions and show how the sentiment towards vaccinations changes over time. Our results highlight an association between differences in network structure and changes to sentiment, demonstrating how network comparison can be used as a tool to guide and enhance the conclusions from sentiment analysis.}





\maketitle

\section{Introduction}\label{sec1}

Online social networks, such as Facebook, Reddit, $\mathbb{X}$ (formerly Twitter), Instagram and LinkedIn, are pervasive in modern society. These platforms can generate immense amounts of data about the interactions of their users, which can be used to inform models of human behaviour. However, the size of these networks means it is often infeasible to study them in their entirety. For example, it was reported that nearly twenty thousand tweets were sent per second on Twitter (now $\mathbb{X}$) during the Qatar 2022 World Cup \citep{musk2022}. Certain subnetworks---communities within these large networks---may also exhibit local behaviour that is different from the global behaviour of the network. As such, it is important when developing models to be able to distinguish between patterns of individual user behaviour and the aggregate statistical properties of such behaviours.

In this work, we apply techniques from network science to study how people communicate online about vaccination. The World Health Organization named Vaccine hesitancy, i.e.\ the reluctance or refusal to vaccinate despite the availability of vaccines, as one of the 10 major threats to global health in 2019 \citep{who2019}. Vaccination is a fundamental tool in the prevention of the spread of infectious diseases, and resistance to vaccination has led to outbreaks of controlled diseases like measles \citep{omer2009vaccine}, putting in danger people who, due to medical conditions, rely on high levels of herd immunity to avoid contracting these infections. Since vaccine ``nonconformers'' \citep{brunson2013impact}---parents who do not vaccinate their children or do so with a subset of the recommended vaccines---tend to be proximately located \citep{brunson2013impact}, the refusal of the MMR (measles, mumps and rubella) vaccination in certain communities reaches values as high as 25\% \citep{omer2009vaccine}.

As might be expected for a health-related topic, vaccination gathers a great deal of traction in online discussions; it has historically been a controversial topic, and remains a matter of considerable debate \citep{blume2006anti}. \citet{brunson2013impact} studied the influence of social and information networks on the vaccination decision and found that the influence of family, friends and medical specialists outweighs that of information sources such as the internet or mass media. However, they also found that ``nonconformers'' seek these information sources significantly more than their ``conformer'' counterparts. A study by \citet{davies2002antivaccination} showed that when searching for keywords ``vaccination'' and ``immunisation'', almost half of the results in the most popular search engines are anti-vaccination. This creates a feedback loop for ``nonconformers'', who seek more information, and thus encounter further information that is anti-vaccination. 

The issue of vaccine hesitancy took on a particular urgency when vaccines against SARS-CoV-2 became available in late 2020, since the COVID-19 pandemic posed not only a health threat but also a threat to social, economic and general well-being worldwide. These threats were particularly intense during mass vaccination campaigns in developed countries in early 2021 and are still ongoing, but greatly reduced by the success of these campaigns. Despite evidence of the efficacy of the vaccines developed against the virus \citep{pouwels2021effect,tregoning2021progress,singanayagam2022community,garciabeltran2022mrna}, concerns were raised about the safety of a vaccine that has been so rapidly developed and tested \citep{deroo2020planning}, particularly in light of reports linking the Astra Zeneca vaccine to rare blood clots~\citep{wisen2021covid19}. Although early studies demonstrated that a high percentage of the population was planning to refuse a vaccine against SARS-CoV-2 \citep{peretti2020future}, rollout was very successful, particularly in the United Kingdom (UK), where it has been reported (at the time of writing this manuscript) that over 50 million second doses and over 40 million boosters have been administered, corresponding to uptakes of 88\% and 70\%, respectively amongst those eligible \citep{govukvaccs}. However, research suggests there is still a high level of vaccine hesitancy within specific communities, such as minority ethnic groups \citep{robertson2021predictors} and those from deprived areas \citep{opensafely2021trends}. Whilst barriers to the uptake of COVID-19 vaccines are still being studied \citep{denford2022exploration}, it is apparent that the motivations for hesitancy are often complex and ever changing \citep{razai2021covid19}.

Social media is a rich source of information about reasons for vaccine hesitancy. The bulk of research in this area is focused on the most prominent and general-interest social media: Facebook \citep{hoffman2019its,schmidt2018social}, Twitter \citep{bello2017detecting,radzikowski2016measles,love2013twitter,addawood2018usage} and Reddit \citep{jang2019social}. Online chat forums are a medium often overlooked by the research community studying vaccination discussions, despite a study by \citet{campbell2017changing} showing that a significant proportion of participants used discussion forums to find out more about immunisation. The same study found that parents who searched for information online were significantly more likely to encounter information that would make them question the decision to vaccinate their offspring, but this likelihood was higher among parents who used discussion forums than those who used Facebook or Twitter. Finally, \citet{rier2007impact} highlights the research potential of online support groups embedded in internet discussion forums, calling them ``natural focus groups" that reflect ``segments of public opinion".

The focus of our analysis in this paper is Mumsnet \citep{mumsnet}, a UK website ``by parents for parents". This website is primarily a place where parents can seek advice on numerous topics, with articles written by other parents, and also offers a forum functionality that allows users to discuss these same topics. Although there are other websites more focused on health issues and therefore vaccination, Mumsnet generates more discussion content regarding vaccination and is more likely to contain a wide range of opinions \citep{skea2008avoiding}. The forum content in Mumsnet is public; anyone can register to comment on any discussion (which appears under a ``username"---an alias that can be used to preserve anonymity) and an account is not required to read or search for past discussions. Prior to July 2019, it contained a subforum dedicated to discussing and asking advice about anything related to vaccination. This was subsequently closed and posts within this subforum were merged into the general health subforum. With the onset of COVID-19, a subforum dedicated to discussing the pandemic was created, in which a substantial body of discussion about COVID-19 vaccines can be found.

One of the most prevalent techniques for examining vaccinations discussions in social media is \emph{sentiment analysis}, that is, identifying and labeling the opinions of social media users about vaccination and applying quantitative or qualitative methods in order to extract knowledge about those discussions. It is a technique often applied to Twitter data (see, for example, \citep{salathe2011assessing,salathe2013dynamics,bello2017detecting,blankenship2018sentiment,muller2019crowdbreaks,muller2020addressing,piedrahita2021vaccine,scannell2021covid}. Sentiment analysis has also been previously applied to study vaccine sentiment in Mumsnet \citep{skeppstedt2017automatic}.

The majority of approaches using sentiment analysis focus on hand labeling a subset of data (e.g.\ tweets or posts) and building machine learning models to classify the remainder of the data. This is the methodology of \cite{skeppstedt2017automatic} and the author's major conclusion is that automated classification in that context is a very hard task, in part due to their small training corpus. This highlights a major issue with the sentiment analysis approach to studying vaccination discussions: obtaining sentiment labeled data is extremely costly, as it is either time consuming or expensive when using platforms such as Amazon Mechanical Turk \citep{buhrmester2016amazon}. There is also a question concerning the reliability  of automatic labeling methods for unseen data, especially when the context and arguments used in the discussion change over time as is the case for COVID-19 vaccine discussions when compared with other vaccine discussions \citep{muller2020addressing}. Finally, using models trained on Twitter data to label non Twitter data is not advisable. As \cite{skeppstedt2017automatic} point out, Twitter messages are limited to 280 characters so the goal of the message is more easily parsed, in contrast to the more complex discussions in Mumsnet posts that are harder to interpret.

Even with the recent sharp gain in popularity by Large Language Models~\citep{brown2020language, touvron2023llama}, sentiment labeling is still an arduous task as these models need to be fine tuned to be able to perform the labeling task, which requires human supervision either by providing hand labeled examples or through reinforcement learning from human feedback. These models are also often not accessible through commodity hardware as the models contain billions of parameters.

Given the issues affecting analyses that rely solely on sentiment labeling, it is of interest to develop methodologies that can reduce the amount of labeling required, or provide additional forms of insight into online debates. One such methodology is to consider the interactions between users that are participating in the discussion, creating a network where two users are connected if they interacted with each other according to some criteria. This perspective on vaccination discussions opens up the opportunity to use the wealth of research created in the area of network science to extract knowledge from these discussions. It is even possible to combine the two methodologies. For example, \citet{yuan2019examining} combine community detection with sentiment analysis. Network analysis has been applied to study the structure of some online discussion forums \citep{romijn2015discovering,zhongbao2003reply}, but Mumsnet has not been studied as a complex network before.

By representing vaccination discussions as a network of the actors contributing to the discussion, we propose studying the structure of the network as a tool to highlight potential changes in sentiment, leading to the two following research questions:

\textbf{RQ1:} What are the characteristic patterns of interactions (captured as complex network structure) between users in vaccination discussions in Mumsnet and how do these differ between discussions of vaccination in general and discussions of vaccination within the context of COVID-19?

\textbf{RQ2:} Over time, are differences in the network structure associated with changes in the sentiment of vaccination discussions within Mumsnet?

To identify changes and differences in the structure of the network, we use techniques aimed at solving the network comparison problem, which quantify differences and similarities between networks, particularly when their sizes differ but the networks are hypothesized to be related. In particular, we employ NetEmd \citep{wegner2017identifying,silva2022comparing}, a network comparison measure that uses small patterns of connections (graphlets) as structural features of the network. Similarly to \emph{motifs} \citep{milo2002network}, graphlets can be seen as units of functional behaviour encoded in the network, characterizing distinct patterns of behaviour. In particular, when considering phenomena that spread---such as viruses, ideas and behaviours---the detailed structure of graphlets in the network in which spreading occurs has an important effect on the dynamics of the process \citep{House:2009,Ritchie:2014,Ritchie:2016,Guilbeault:2018}. As such, when we observe patterns in the graphlets relating to interactions in online discussions, this will hold information about the manner in which opinions are spread and adopted. By using NetEmd, we can identify changes in the patterns---indicating differences in the network structure---and extract the graphlets responsible for these changes.

The major contributions of this work can be summarized as thus:

\begin{itemize}
    \item We analyse sentiment towards vaccination in Mumsnet over time, with an emphasis on the period before and after the COVID-19 pandemic.
    \item We show how to use network comparison to track topological changes in network structure over time, finding significant differences in the structure of user interaction networks before and after the COVID-19 pandemic.
    \item By performing an observational study, we find an association between differences in network structure and changes in sentiment towards vaccination.
\end{itemize}

\section{Background}

\subsection{Network Terminology}

\textbf{Graphs} are mathematical objects used to represent a complex network. A graph $G = (V(G),E(G))$ is composed of a set of nodes $V(G)$ and a set of edges $E(G) \subseteq V(G) \times V(G)$, where each edge is represented by a pair $(u,v) \in E(G)$ for $u, v \in V(G)$. The graphs we consider in this work are \textbf{undirected}, meaning that the order of the vertices in the pairs does not express direction, so $(u,v)\in E(G)$ implies that $(u,v)\in E(G)$. Graphs are assumed to be labeled, all nodes are assigned consecutive integer numbers starting from $0$ and running to $\lvert V(G)\rvert-1$. The \textbf{size} of a graph is the number of vertices in the graph, written as $\lvert V(G)\rvert$. A size-$k$ graph is a graph with $k$ vertices. The \textbf{density} of a graph is the portion of edges present in the graph over the total potential ones, calculated as $\lvert E(G)\rvert/\binom{\lvert V(G)\rvert}{2}$ in undirected networks.

Graphs can be classified into different types according to constraints or additions to the set of nodes and edges. A graph is called \textbf{simple} if it does not contain multiple edges (two or more edges connecting the same pair of vertices) or self-loops (an edge of the form $(u,u)$ that connects a vertex to itself). In this work, we consider only simple graphs. A graph is called \textbf{connected} if all nodes are reachable from each other. A node $v \in V(G)$ is reachable from a node $u \in V(G)$ if $(u,v) \in E(G)$ or if there is a path between $u$ and $v$, i.e., a sequence of edges of the form: $(u,t_1), (t_1, t_2), \ldots (t_n, v)$. A \textbf{bipartite graph} is a graph whose vertices represent two disjoint groups of entities and there are no edges between members of the same group. Formally, $G = (V(G_1) \cup V(G_2), E(G))$, such that $V(G_1) \cap V(G_2) = \varnothing$ and $E(G) \subseteq V(G_1) \times V(G_2)$, with each edge represented by a pair $(u,v) \in E(G)$ for $u \in V(G_1) \Rightarrow v \in V(G_2)$ or $u \in V(G_2) \Rightarrow v \in V(G_1)$. A \textbf{weighted graph} is a graph whose edges represent additional information, a \textbf{weight}, about the relationship between vertices, beyond the existence of that relationship. The most common example of an edge weight is a numeral attribute representing, for instance, the frequency of the relationship. In the case of a numerical attribute, the set of edges is defined as $E(G) \subseteq V(G)\times V(G) \times \mathbb{R}$ and each edge is represented as a triple $(u,v,w) \in E(G)$ for $u, v \in V(G)$ and $w \in \mathbb{R}$.

The \textbf{neighbourhood} of a vertex $u \in V(G)$ is the set of nodes that $u$ is connected to and defined as $N(u) = \{ v : (v,u) \in E(G) \lor (u,v) \in E(G) \}$. The \textbf{degree} of a vertex is the number of edges it participates in, which is equivalent to the size of the node's neighbourhood, $\lvert N(u)\rvert$. A node is called \textbf{isolated} if it has no connections, i.e., its degree is 0. The \textbf{weighted degree} of a node is the sum of weights of all edges the node participates in.

A \textbf{subgraph} $G_k$ of a graph $G$ is a size-$k$ graph such that $V(G_k) \subseteq V(G)$ and $E(G) \supseteq E(G_k) \subseteq V(G_k) \times V(G_k)$ and is called \textbf{induced} if $\forall u,v \in V(G_k), \; (u,v) \in E(G_k) \leftrightarrow (u,v) \in E(G)$. Two graphs $G$ and $H$ are \textbf{isomorphic}, written as $G\sim H$, if there is a bijection between $V(G)$ and $V(H)$ such that two vertices are adjacent in $G$ if and only if their correspondent vertices in $H$ are adjacent. A \textbf{match} of a graph $H$ in a larger graph $G$ is a set of nodes that induce the respective subgraph $H$. In other words, it is a subgraph $G_k$ of $G$ that is isomorphic to $H$. The frequency of a subgraph $G_k$ is then the number of different matches of $G_k$ in $G$.

\textbf{Orbits} are unique positions of a graph, calculated by partitioning the set of vertices into equivalence classes where two vertices belong to the same class if there is an automorphism that maps one into the other \citep{ribeiro2019survey}.

\begin{figure}
    \centering
    \includegraphics[width=0.95\textwidth]{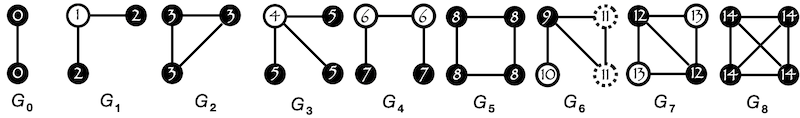}
    \caption{Undirected graphlets of size 2, 3 and 4. Nodes are numbered according to their orbit and nodes with the same number represent the same orbit.}
    \label{fig:graphlets}
\end{figure}

\textbf{Graphlets} are small, connected, non-isomorphic and induced subgraphs \citep{prvzulj2007biological}. The smallest graphlet considered is a single edge, which can be seen as a size-2 subgraph. An undirected edge has a single orbit (three in the directed case) and the frequency of a node in this orbit is equivalent to the degree of the node. Figure~\ref{fig:graphlets} shows the graphlets of size 2, 3 and 4 in undirected networks, alongside the respective orbits. 

A graph where all nodes are connected to each other is called a \textbf{clique}. In other words, it is a graph with density of 1. In Figure~\ref{fig:graphlets}, corresponds to undirected graphs $G_2$ and $G_8$. 

A graph $G$ is called \textbf{bipartite} where we can separate the vertex set into two disjoint subsets so that $V(G) = V_1(G) \cup V_2(G)$ and $V_1(G) \cap V_2(G) = \varnothing$ and edges only exist linking elements of different vertex sets, i.e., $E(G)\subseteq (V_1(G)\times V_2(G)) \cup (V_2(G) \times V_1(G))$.

\subsection{NetEmd}

NetEmd \citep{wegner2017identifying} is a network comparison measure that relies on structural features of the network, mainly the distribution of orbit frequencies. The core idea is formalizing the intuition that the shape of the degree distribution is indicative of the network's generation mechanisms, for instance, a network with a power law degree distribution is generated by a process distinct from a network with a uniform degree distribution. As the graphlet degree vector is a generalization of the degree distribution for graphlets of size $k \geq 3$, the shapes of the distributions of each orbit also carry information about the topology of the network. Note that because the shape of a distribution is invariant under linear transformations such as translations, using the shape as the focus of the comparison is well-suited to comparing networks of different sizes and densities. \citet{wegner2017identifying} define a measure of similarity between distributions $p$ and $q$, with non-zero and finite variances, using the Earth Mover's Distance (EMD) \citep{rubner1998metric}:

\begin{equation*}\label{eq:emd}
\mathrm{EMD}^{*}(p,q) = \text{inf}_{c \in \mathbb{R}} (\mathrm{EMD}(\tilde{p}(\cdot + c), \tilde{q}(\cdot))),
\end{equation*}
where $\tilde{p}$ and $\tilde{q}$ are the distributions resulting of scaling $p$ and $q$ to variance 1. Any distance metric $d$ can be used to generate $d^*$; EMD was chosen here since it has been shown to be an appropriate metric to compare shapes of distribution in domains such as information retrieval and it produced better results than other distance metrics, like Kolmogorov ($\ell_\infty$) or Manhattan ($\ell_1$) distances.

Given two networks $G$ and $H$ and a set of $m$ orbits $\mathcal{O} = \{o_1, o_2, \ldots, o_m\}$, the \emph{NetEmd} measure is defined as:

\begin{equation}
\label{eq:netemd}
\mathrm{NetEmd}_{\mathcal{O}}(G,H) = \frac{1}{m} \sum\limits_{i=1}^{m} \mathrm{EMD}^*(p_{o_i}(G),p_{o_i}(H)),
\end{equation}
where $p_{o_i}(G)$ and $p_{o_i}(H)$ are the distributions of orbit $i$ in graphs $G$ and $H$, respectively. Note that $\mathcal{O}$ may be replaced by any set of network features; in this work we focus on orbits only.

\cite{silva2022comparing} propose an extension of NetEmd by noting that different orbits do not represent independent degrees of freedom and are expected to be constrained by the requirement of combinatorial consistency of the underlying full graph. These combinatorial constraints lead to added noise during the EMD calculation, thus the authors propose using linear techniques of principal component analysis (PCA) and independent component analysis (ICA) as a noise reduction mechanism. The idea is to project the orbit frequencies to a lower dimension, training the dimension reduction model to lose minimal information while removing noise, project the reduced features to the original dimension and apply the NetEmd equation to the reconstructed frequencies.

\subsection{Mumsnet}

The Mumsnet forum has been the subject of a number of vaccination studies \citep{skeppstedt2017automatic,skeppstedt2018vaccine,ford2018use} with the MMR (measles, mumps and rubella) vaccine being a particular focus \citep{skea2008avoiding,weller2016content}. 

\cite{skeppstedt2017automatic,skeppstedt2018vaccine} explored 5 threads about general vaccination of children, manually labeling a dataset of 1190 posts as for, against or undecided about vaccination. The authors then built a machine learning model to measure the performance of an automatic classifier on this labeled dataset. The machine learning model chosen was a linear support vector machine \citep{cortes1995support} and the training data was obtained by preprocessing each post into a list of tokens and n-grams, i.e., sequences of $n$ consecutive words from each post (the chosen values for $n$ were 2, 3 and 4). The authors found among the manually annotated posts an even distribution of sentiment, with 38\% for and 41\% against. The machine learning experiment yielded a model capable of identifying the sentiment of this corpus with an $F_1$-score of 0.44, which lead the authors to conclude that their small training corpus makes classification a very hard task.

\cite{ford2018use} used Mumsnet as a platform for a cross-sectional study of influenza and pertussis vaccination uptake during pregnancy. The study's objective was to learn whether information obtained from online social networks (Facebook, Twitter, etc.) influenced the decision to have these two particular vaccinations. They found that women who used social networks to obtain information were less likely to receive the pertussis vaccination, but this was not the case for the influenza vaccination.

\cite{skea2008avoiding} examined two threads about the MMR vaccine from the perspective of parental perception of herd immunity, and the duality between parents not wanting to harm their child versus wishing to avoid harm to others. The authors suggested that vaccine promotional material should ``include explanations of herd immunity'' as it may influence parents' decisions. \cite{weller2016content} analyzed posts and threads related to MMR, finding that by 2010 the majority of the forum was in favour of MMR rather than against it.

\section{Methods}

\subsection{Data Collection}

The Mumsnet forum is organized in categories, which are further divided into more specialized topics, called subforums, within those categories. Users create \emph{threads} on these subforums by typing a \emph{thread title} and some text, called the \emph{original post}. 
Other users can reply to the original post by typing in a text box; these replies, called \emph{posts}, are displayed in chronological order of submission. One of the categories of the forum is \textit{health} and within it, there was a subforum about vaccination, which was closed down in July 2019, and another subforum about the COVID-19 pandemic, created in February 2020. We collect two datasets from Mumsnet, one from each of these subforums.

Prior to the closure of the vaccination subforum, we deployed a web scraper to acquire all threads and posts publicly available in the subforum, spanning back 10 years. Following the closure of the forum, we set the start date for our data collection to 1 July 2009, with the last post on this forum dating 28 June 2019. In this time period, there were 1128 threads created, with a total of 31,757 posts, written by 5715 users. As a shorthand, this dataset will be referred to as the \emph{vaccine} or \emph{vaccination} dataset.

The COVID-19 subforum contains threads about any topic related to the pandemic, so we filtered these according to their title. We included all titles that contained keywords related to vaccination and immunization (essentially a regular expression matching \textit{vaccin*}, \textit{immun*} or \textit{*vax*}), as well as COVID-19 vaccine specific terms, for instance, booster or the names of the vaccine manufacturers. There were only 20 threads before 1 November 2020, compared to 81 in November 2020 alone, so we set this as the start date for data analysis. We finalized the data collection on 2 January 2022 and the final dataset only includes posts dated before 2022. In this time period between 1 November 2020 and 31 December 2021, there were 4322 threads created, with a total of 223,625 posts, written by 25,989 users. Again as a shorthand, we refer to this dataset as the \emph{COVID-19} dataset and every time we refer to the COVID-19 forum, we mean vaccination discussions in particular within this forum.

\subsection{Data Segmentation by Time}

In order to study how discussions vary over time in Mumsnet, we segment the datasets into new and smaller datasets that span shorter  periods. One way to accomplish this goal is to create disjoint time intervals so that there is no overlap between the smaller datasets. To do this, we would need a rule for allocation of posts to time periods. Consider the follow situation: a thread is created on 31 January; half the posts within that thread are submitted on that day and the other half are posted throughout February; and time periods are delineated by the first day of each month. One way to assign data to each time frame is by the date of the thread, but this means that the February dataset does not encompass the discussion that took place in that thread throughout February. Alternatively, half the posts within that thread are assigned to the January dataset and the other half to February. In this case, we are separating discussion that took place within a similar context by imposing an arbitrary boundary; it could be argued that a post from 31 January is more similar to a post from 1 February than to a post from 1 January.

To solve this, we create a rolling window of data, akin to a moving average. We assign posts to data slices according to their post date and ensure that related posts on the edge of the time split are assigned to the same data slice at least once. This method of temporally separating the data is controlled by two parameters, the \emph{time span}, i.e., how much time each window covers, and the \emph{time jump}, i.e., the amount of time between consecutive windows. The \emph{time overlap} between consecutive windows, i.e., the amount of time in common between consecutive windows, is calculated by subtracting the time jump from the time span.

\subsection{Network Creation}
\label{sec:net_creation}

The way discussion is carried out in a forum is different from other social media platforms like Facebook, Twitter or Reddit, in the sense that users' interactions with threads and messages appear sequentially ordered by time, regardless of relevance to the discussion or tagging features used. This structure lends itself to the creation of a bipartite network, as two disjoint sets of entities naturally emerge: threads and users, with a user $u$ connected to a thread $t$ if $u$ wrote a post in $t$. A common technique for analyzing bipartite networks is to create a projection of one of the partitions, for example when one of the modes in the data holds particular interest over the other \citep{borgatti1997network}. The most common way to achieve this projection is assuming that if two nodes in one partition are connected to the same node in the other partition, then they should be connected in the projection. In our case, since we are interested in studying patterns of connections between users, we create a user projection network, where users are connected if they posted in the same thread. Formally, if we let $\mathbf{X} = [x_{it}]$ be the bipartite adjacency matrix, with row labels corresponding to users and column labels corresponding to threads, then the user projection network adjacency matrix elements can be obtained as outlined by \citet{borgatti1997network} as
\begin{equation}
\label{eq:1modeproj}
a_{ij} = \sum\limits_{t \in \mathcal{T}} x_{it} x_{jt},
\end{equation}
where $\mathcal{T}$ is the set of threads in the bipartite network and $x_{it}$ represents the number of posts of user $i$ in thread $t$. In matrix notation, this can be simplified as $\mathbf{A} = [a_{ij}] = \mathbf{X}\mathbf{X}^T$. This transformation produces a weighted undirected network, but it is not representative of the reality in forums, as it works under the assumption that all users in a thread read each other's posts. Inspired by \cite{newman2001scientific}, we \textit{weight} Eq.~\ref{eq:1modeproj} to take into account the number of users and number of posts in a thread, similar to collaboration networks where a connection between two researchers is weaker when they publish a paper with a lot of co-authors and stronger when they publish a paper together or with a small number of other co-authors. The strength of the connection is also proportional to the quantity of papers where the researchers are co-authors, with more papers indicating a stronger connection. In our case, two users have a weak connection if they both post in a thread with many posts and a strong one when they participate in multiple threads together. The formula to calculate the projection network adjacency matrix then becomes the following:
\begin{equation}
\label{eq:1modeweightedproj}
a_{ij} = \sum\limits_{t \in \mathcal{T}} \frac{x_{it} x_{jt}}{(\upsilon_t - 1)\, \phi_t},
\end{equation}
where $\upsilon_t$ is the number of users in thread $t$ and $\phi_t$ is the number of posts in thread $t$. We define a minimum weight for connections between users, below which we assume that the users had insignificant influence over each other. This threshold was chosen as the maximum value that keeps the network connected, i.e., with no isolated nodes.

\subsection{Network Comparison}

The networks we create from Mumsnet data are a means of representing user posting behaviour and how users interact with each other within Mumsnet. A change in network structure therefore indicates a change in user behaviour, which in turn may reflect changes in how people are discussing or organising around a topic.

To identify differences in the structure of Mumsnet networks, we use NetEmd \citep{wegner2017identifying} to compare each pair of networks. As these networks share the same generation mechanism, we expect NetEmd to output a high degree of similarity between them; deviations from this similarity indicate that there is a difference in structure. Upon calculating the comparison, we can detect differences in structure by looking at the distance between networks from consecutive data slices. If this distance is larger than the median distance between any pair of networks, then we consider that there is a significant difference in the structure of the network. Due to overlapping time frames, differences in the network structure may also take two or three time steps to become apparent. 

As graphlets can be seen as small units within a network whose patterns of connections encode real world functionality, we can interpret the network comparison results by determining which orbits are responsible for the largest difference, and inspecting these.

\subsection{Sentiment Labelling}

Returning to the work of \cite{skeppstedt2017automatic} and their approach to automatically labeling Mumsnet posts according to their stance on vaccination, we note that the authors highlight the difficulty of this task, particularly in comparison to finding stance in tweets, as posts are usually longer than tweets, given that they are not restricted to 280 characters, and therefore contain longer and more elaborate discussions. The approach taken by the authors yielded a machine learning model that struggles to learn how to classify such an elaborate corpus, achieving an $F_1$ score of 0.44.

Starting from \cite{skeppstedt2017automatic}'s approach, we attempted to improve the classification performance of a machine learning algorithm in the same labeled dataset, with the intent of ultimately applying it to the larger corpus we consider in our work. We found that using a Random Forest \citep{breiman2001random} improved the classification performance slightly (up to a $F_1$ score of 0.52). Other approaches using the Valance Aware Dictionary and Sentiment Reasoner \citep{hutto2014vader}, the Na\"{i}ve Bayes classifier of \cite{salathe2011assessing} and a FastText model \citep{joulin2017bag} trained on the Crowdbreaks platform \citep{muller2019crowdbreaks} all performed worse than adapting the code of \cite{skeppstedt2017automatic} to work with Random Forests.

The poor performance of automatic labeling of sentiment in the training dataset of Mumsnet posts disqualifies this methodology from being applied to the larger corpus of posts, as the labeling would be unreliable. Another reason to avoid using a machine learning model trained on Skeppstedt et al.'s data is a \emph{concept drift} \citep{widmer1996learning}. This captures the idea that models trained on one snapshot of time may struggle to generalize for data outside of that time period, particularly in language based models trained on internet data, where context changes over time. \cite{muller2019crowdbreaks} showed the existence of concept drift in vaccination discussions by analyzing a Twitter dataset in the second half of 2018 and at the beginning of the COVID-19 pandemic \citep{muller2020addressing}. The issue of concept drift is particularly relevant when comparing vaccination discussions before and after the COVID-19 pandemic. Changes in the vocabulary, arguments, context and the general discourse in these discussions lead to an increasing likelihood that a model trained on pre-pandemic data will not be able to accurately label a post-pandemic corpus.

The issues we identified with automatic labeling of Mumsnet posts led to the decision to manually label the two datasets we acquired from Mumsnet, to create a ground truth against which we can compare the methodology we develop. 
We label threads according to the title and the original post. We classified threads using three labels: \emph{positive}, \emph{neutral} or \emph{negative}. We do not consider negative sentiment to be equivalent to an anti-vaccination position, as it is possible to express negativity towards a vaccine or a vaccination schedule while simultaneously adhering to the policy that recommends taking it. An example of this is starting a discussion about suffering from side effects of a vaccine, without asking for help on how to overcome them, expressing regret over taking a vaccine or doubting the efficacy of a particular vaccine. The following is a non-exhaustive list of common arguments we marked as negative towards vaccination:

\begin{itemize}
    \item claiming to have refused to be vaccinated and to not plan on being in future;
    \item wishing to vaccinate on a different schedule, delaying recommended or mandatory vaccines;
    \item showing lack of trust in or avoiding altogether vaccine ``cocktails'' (a disparaging term for combined vaccines, most commonly MMR);
    \item blaming vaccines for illness or expressing regret about taking the vaccine;
    \item questioning the efficacy of the vaccine or the safety of ingredients within the vaccine (excluding allergic reactions);
    \item supporting known opponents of vaccination (e.g., Andrew Wakefield) or anti-vaccination documentaries (e.g., ``The Greater Good").
\end{itemize}

The neutral label captures two types of posts: those asking for information without expressing a judgment of value towards the vaccine itself; or those with conflicting positive and negative views about vaccination (for example, planning to vaccinate on schedule but mentioning being afraid or anxious). Finally, we marked threads as positive if they contained arguments that were supportive of vaccination, for example detailing its benefits, encouraging calls for more vaccines, explaining consequences of the disease that can be prevented through vaccination or mentioning someone going out of their way to get a vaccine, for example privately.

The manual labeling process was completed by two annotators (MEPS and RS), who split the annotation load equally, with each annotating half the threads of each dataset. The threads were assigned uniformly at random to each annotator. Prior to annotating, the two annotators labeled 1\% of the COVID-19 dataset together (40 threads, chosen uniformly at random), showing high reliability with a Cohen's Kappa score of 0.84 (90\% agreement). The annotators also discussed the results to settle discrepancies and agree on the mutual strategy for the remainder of the data.

In order to study the distribution of sentiment within threads of the COVID-19 dataset, we sampled 129 threads (about 3.0\% of the number of threads) which contained 6898 posts (about 3.1\% of the number of posts), chosen uniformly at random such that they represent at least 1\% of the threads and posts in each data slice. The sentiment in all these 6898 posts was labeled by a single annotator (MEPS) using the same criteria used to label the threads.

\subsection{Discordance Index}

Under the hypothesis that the proportion of posts of each sentiment within a thread alone is not sufficient to capture the extent of debate and disagreement, we propose an alternative metric to measure the level of disagreement, which we call the \emph{discordance index}. This metric measures disagreement at a local level by capturing how often sentiment changes in small groups of consecutive posts and using the number of sentiment changes as a proxy for the level of discordance, weighting changes from positive to negative (or vice-versa) higher than changes involving neutral sentiment.

An example that motivates this metric is the following: consider a thread with $N$ posts, where the first $N/2$ posts are of negative sentiment and the last $N/2$ are positive. By considering only the proportion of positive to negative posts, this thread would appear to contain a large amount of disagreement between users as half the posts express a diametrically opposed view to the other. However, by inspecting how the sentiment is distributed over the thread, we actually find that most users are agreeing with each other, with the exception of the shifting point from negative to positive. On the other hand, if a thread with $N$ posts has $N/2$ negative posts interleaved with $N/2$ positive posts, then it is much more likely that users are constantly disagreeing with each other throughout the thread. In both cases the proportion of negative to positive posts is the same, but the way discussion is being carried out is very different. 

Our proposed metric \emph{discordance index} attempts to capture this distinction of local variety of sentiment within a thread. The way it works is by considering a moving window of $D$ posts, such that in a thread with $N$ posts there are $N-D+1$ windows of $D$ posts. For each window, we compare every unique pair of posts within that window ($D(D-1)/2$ comparisons). If one post is positive and the other is negative, we assign a distance of 2; if a post is neutral and the other is positive or negative, we assign a distance of 1; finally, if the posts are of the same sentiment, we assign a distance of 0. The discordance within a window is the sum of distances between each pair of posts; the discordance of a thread is the sum of discordances of all windows, divided by the maximum discordance for that window size. The discordance index of a thread is the average discordance across multiple window lengths, which we vary between 2 and 5.

\section{Results}

\subsection{Data Granularity}

We start by identifying the appropriate window overlaps to split each dataset over time. Although answering the research questions we posed in the introduction should not depend on these parameters, they are responsible for controlling how much change and variability are in each data slice. For instance, picking too long a window span may lead to a uniform sentiment throughout the whole dataset, whereas too small a window span could lead to so much variability between consecutive time windows  that any difference becomes meaningless.

\begin{figure}
    \centering
    \begin{subfigure}{0.5\textwidth}
    \centering
    \includegraphics[width=\textwidth]{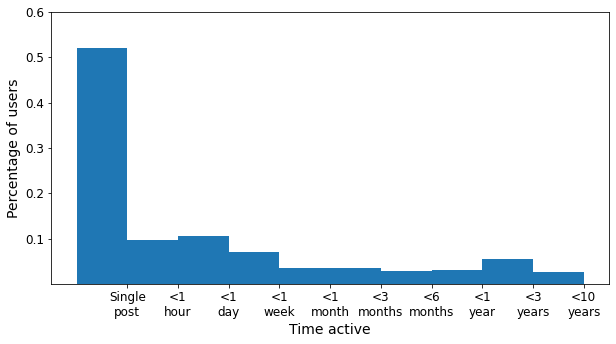}
        \caption{Vaccine forum -- user active times}
        \label{fig:vaccine_user_times}
    \end{subfigure}%
    \begin{subfigure}{0.5\textwidth}
    \centering
    \includegraphics[width=\textwidth]{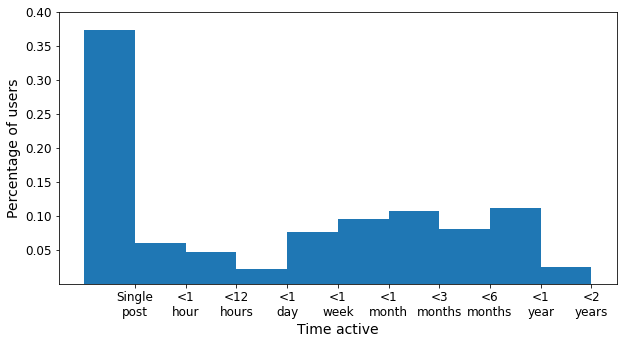}
        \caption{COVID-19 forum -- user active times}
        \label{fig:covid_user_times}
    \end{subfigure}
    \begin{subfigure}{0.5\textwidth}
    \centering
    \includegraphics[width=\textwidth]{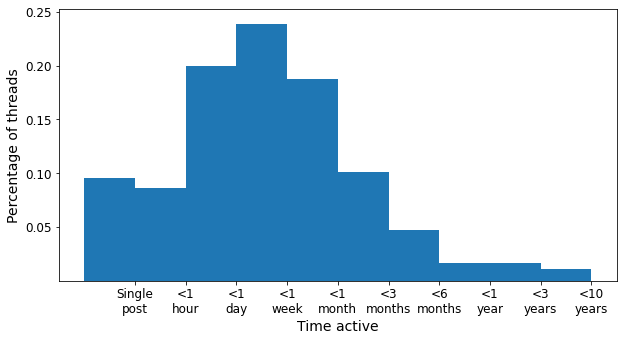}
        \caption{Vaccine forum -- thread active times}
        \label{fig:vaccine_thread_times}
    \end{subfigure}%
    \begin{subfigure}{0.5\textwidth}
    \centering
    \includegraphics[width=\textwidth]{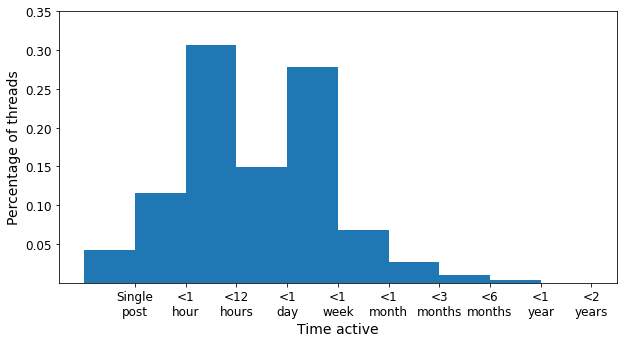}
        \caption{COVID-19 forum -- thread active times}
        \label{fig:covid_thread_times}
    \end{subfigure}%
    \caption{Empirical distributions of user and thread active times. User active times: distribution of how long each user remains active in each forum, obtained by taking the time difference between the date of their first and last posts. Thread active times: distribution of how long each thread remains active in each forum, obtained by taking the time difference between the date of their first and last posts.}
    \label{fig:active_times}
\end{figure}

The first result we use to inform the decision is looking at how long each user spends on the forum and how long each thread remains active (Figure~\ref{fig:active_times}). These figures show that users behave differently in the two subforums. In the vaccine dataset, we find that 52\% of users post only once and close to 10\% of threads get no replies. This contrasts with the COVID-19 dataset, where 37\% of users have only one post and 4\% of threads did not get any replies. 

The results show that user active times---the period between a user's first and last post---are shorter in the vaccine forum than the COVID-19 forum: 80\% of users in the vaccine forum are active for less than a week; in the COVID-19 forum, if we exclude users with a single post, only 20\% of users are active for less than a week, and 42\% of users are active for longer than this. This may be because the pandemic was an extremely disruptive event simultaneously affecting people worldwide, whereas the issues brought up in the vaccine subforum are more likely to be short lived or affect individuals. Despite this, results highlight that there is a core group of users in the vaccine forum with long term engagement: 8\% of users were active on this forum for longer than a year and close to 3\% were active for more than 3 years.


Inspecting the thread active times in the COVID-19 forum highlights how fast information changes during a global pandemic, particularly in the modern age where information is easily disseminated and accessible. Results show that 42\% of threads last less than 12 hours (46\% if taking into account posts without replies); close to 28\% of threads last less than a week but longer than a day and only 11\% of threads last longer than a week. This is likely because as new information about the vaccines was being released, older discussions quickly lost their relevance. This contrasts with how discussions were being held in the vaccine forum, with only 62\% of threads lasting less than a week and 1.1\% of discussions lasting longer than 3 years, compared with only 1.3\% of COVID-19 discussions lasting longer than 3 months.

These two sets of figures lead us to conclude that we should favour a longer time span for each data slice from the vaccine dataset, because many users post for a short amount of time but their influence can be long lasting due to a substantial number of discussions lasting multiple months. On the other hand, the time span of data slices from the COVID-19 dataset should be smaller, as threads tend to last a short time and users show signs of repeated engagement with the forum.

Based on these observations, we narrow down the candidates for window span and window jump and decide to use a combination thereof by inspecting how the number of posts, users and threads varies over time (several combinations are visualized in Figures~\ref{fig:vaccine_data_granularity} and~\ref{fig:covid_data_granularity}, for the vaccine and COVID-19 dataset respectively). For the vaccine dataset, we set the time windows to have a 4 month span with 2 month time jump, leading to a 2 month overlap between consecutive data slices. For the COVID-19 dataset, we pick a 1 month span with a 2 week time jump, leading to a 2 week overlap between consecutive data slices.

\subsection{Temporal Network Comparison}

Upon splitting the datasets and creating a network using the process described in Section~\ref{sec:net_creation} for each time window, we obtain two datasets of networks. The networks in each dataset correspond to users connected through their interactions in a medium that does not change and has been created using the same the methodology, making it reasonable to assume that the networks have a similar structure. 
If the networks are dissimilar according to the NetEmd network comparison method, this indicates that the way that users are interacting within the forum is also different.

  

In both datasets, we observe that the number of nodes and the average degree varies significantly over time. In the vaccine dataset, the number of nodes varies between 101 and 469 and the average degree between 4.6 and 19.7. In the COVID-19 dataset, the number of nodes fluctuates between 1873 and 5743 and the average degree between 25.7 and 84.8. Experiments by \cite{silva2022comparing} indicate that the best performance on real world datasets of undirected networks with this disparity in the number of nodes and average degree is achieved by using size 4 graphlets. As indicated by their experimental setup, we use NetEmd with PCA and 90\% explained variance, although we find that, in practice, $>90\%$ of the structure differences signaled by this parametrization overlap with the ones found by the original NetEmd.

The symmetric distance matrix returned by NetEmd containing the distances between each pair of networks can be visualized with a heat map. This visualization technique helps to highlight clusters of networks that are similar and quickly identify networks that differ substantially from their predecessors and successors. The heat map for the vaccine dataset is shown in Figure~\ref{fig:vaccine_dists} and for the COVID-19 dataset in Figure~\ref{fig:covid_dists}. On the axis of each heat map, we include the \emph{start date} of the time window each network is built from. For example, in the vaccine dataset, the first network, labeled as ``07-2009", has data spanning from 1 July 2009 to 31 October 2009; in the COVID-19 dataset, the first network, labeled as ``11-2020", has data from 1 November 2020 to 30 November 2020, the second network (not labeled) from 15 November 2020 to 15 December 2020, the third (labeled ``12-2020") from 1 December 2020 to 31 December 2020 and so on.

\subsubsection{Vaccination forum}

\begin{figure}
    \centering
    \includegraphics[width=0.95\textwidth]{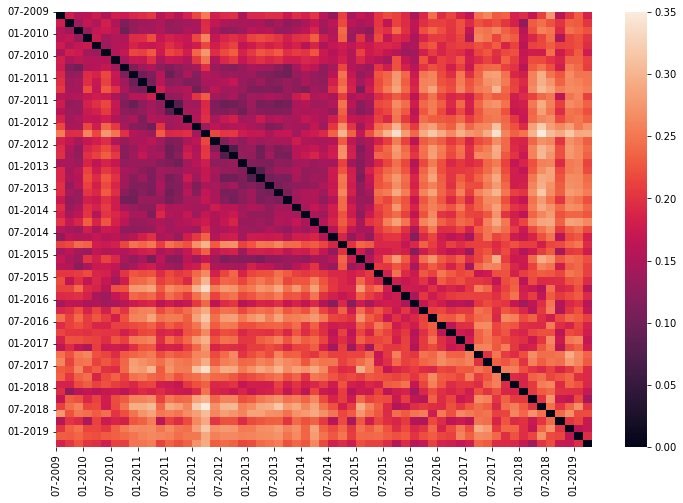}
    \caption{Heat map of network distances between each pair of networks in the Mumsnet vaccine dataset. Distances are calculated using $\mathrm{PCA\_NetEmd}$ \citep{silva2022comparing} with size 4 graphlets and 90\% explained variance.}
    \label{fig:vaccine_dists}
\end{figure}

Inspection of Figure~\ref{fig:vaccine_dists} reveals two broad groups of networks -- before and after May 2015. Prior to this date, the heat map shows two networks that are substantially different to neighbouring time slices, one in March 2012 and the other in September 2014; but the remaining networks before May 2015 possess a high degree of similarity, as indicated by the darker hue. After May 2015, the network structure changes more frequently, as shown by the lighter hue colouring the distances between these networks. Although they are nonetheless more similar to each than to networks before May 2015, the distances among elements of this post May 2015 cluster are, on average, higher than between the networks that compose the pre-May 2015 cluster. It is also of note that the transition between these clusters is gradual, starting with the network from May 2014.

Some of these differences in network structure occur contemporaneously with controversies surrounding vaccines. By inspecting post content in the data slice where we detect network differences, we found evidence that these news stories and controversies penetrate discussions in the forum. For instance, the differences observed in the network starting in May 2014, coincide with the publication of a study \citep{hooker2014measles}, where the connection between the MMR vaccine and autism is investigated. Another example is the network starting in May 2015, during which a child whose parents opposed vaccination died of diphtheria in Spain~\citep{diphtheria}. There is also evidence changes in structure coincide with controversial topics introduced by users. 
For example the difference between the networks of March 2012 and May 2012 could be related to a discussion regarding the safety of aluminium in vaccines.

To quantify the changes we observe in the heatmap, we compare the distance value between networks of consecutive time windows with the median of the values in the upper triangle of the distance matrix. We also compare this median against the distance value between networks from time windows that are 2 jumps apart, pairs of networks with no overlapping data but spanning a continuous time frame. Our measure for significance of difference between consecutive networks is whether their distance is greater than the median of distances between all pairs of networks. Prior to May 2015, we only find significant differences in the time windows we discussed previously: March 2012 to May 2012 and May 2014 to September 2014. After May 2015, we detect 8 significant differences between networks from consecutive time slices and 11 when comparing with 2 jumps apart.

The NetEmd measure between two networks is calculated using the average distance for each orbit, allowing us to pinpoint which graphlets are driving changes in the network structure by inspecting the distances between distributions of each orbit. The heat maps of network distances for each individual orbit are shown in the Appendix, Figure~\ref{fig:vaccine_dists_allorbs}. We focus on two time periods as an example:
\begin{itemize}
    \item The difference between the networks starting in March and May 2012 is primarily connected with the distributions of orbits 1, 6 and 9 (refer to Figure~\ref{fig:graphlets}). The common element between these 3 orbits is that within their graphlet, they are connected to a node with no other connections.
    \item The distinction between the clusters before and after May 2015 is largely connected with differences in orbits 0 (the degree distribution), 3 and 14, which are all orbits part of cliques.
\end{itemize}

\subsubsection{COVID-19 forum}
\label{sec:covid_net_struct}

\begin{figure}[t]
    \centering
    \includegraphics[width=0.95\textwidth]{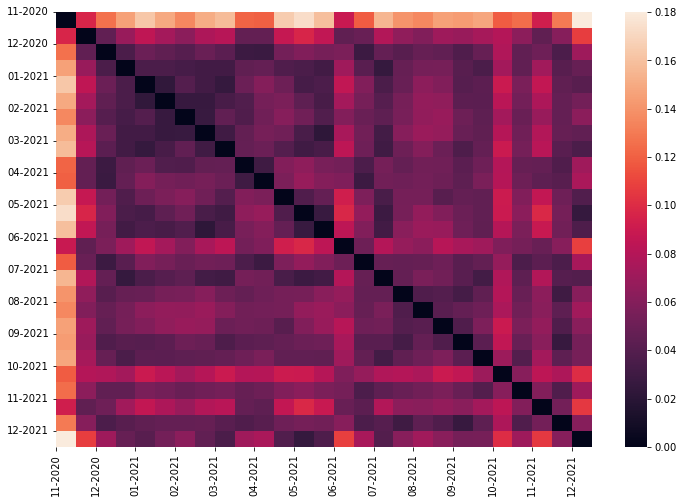}
    \caption{Heat map of network distances between each pair of networks in the Mumsnet COVID-19 dataset. Distances are calculated using $PCA\_NetEmd$ \citep{silva2022comparing} with size 4 graphlets and 90\% explained variance.}
    \label{fig:covid_dists}
\end{figure}

Comparing networks from the COVID-19 dataset yields an altogether different pattern of structures, as shown in Figure~\ref{fig:covid_dists}. The heat map of distances is dominated by darker hues, meaning that the majority of networks are very similar. The exception to this is the first network, which spans the month of November 2020 and is the only network in this dataset that precedes the announcement of the first COVID-19 vaccine being approved~\citep{pfizerapproval}. Recall that the first vaccine to be approved in the UK was the Pfizer vaccine on 2 December 2020, so the second network in this dataset, starting on 15 November 2020, already includes this period. This second network also shows a high degree of dissimilarity when compared with the remaining networks in this dataset, but not to the same extent as the first one.

Two other periods of time stand out when analyzing the heat map: the networks starting on 1 June 2021 and 1 October 2021. The former coincides with the Delta variant becoming the dominant strain of COVID-19 in the UK~\citep{torjesen2021covid}, which was thought to be more resistant to COVID-19 vaccines than previous variants, in particular when only a single dose had been administered, at a time when approximately 50\% of the population had received the second dose. The network with a start date in October 2021 coincides with the beginning of the booster roll out in the UK~\citep{iacobucci2021covid}, which seems to be the leading topic of discussion during this time as ``booster" was the sixth most common word among all posts in October 2021 (when removing stop words, after ``COVID", ``vaccine", ``people", ``think" and ``vaccinated").

Applying the same analysis as before--comparing the distance between networks from consecutive and 2 jump apart time windows against the median distance between all pairs of networks--reveals new points in time where there is a significant difference in network structure. In addition to the previous three time windows, we also find networks with a start date of 15 March 2021 and 1 April 2021 to be significantly different from their neighbouring time periods. These two networks coincide with the blood clot controversy surrounding the Astra Zeneca vaccine~\citep{bloodclot}. Finally, we also find a significant difference between the network starting on 1 November 2021 and the network starting on 1 December 2021. This difference coincides with the emergence of the omicron variant~\citep{elliott2022rapid} and the government's push for more boosters to increase protection against the new variant.

When inspecting the differences between the distributions of each orbit, visualized in Figure~\ref{fig:covid_dists_allorbs}, we find that the network of 1 November 2020 is significantly different from the other networks in every orbit except 8, where this network is similar to the following two networks (15 November and 1 December 2020). Similarly, the difference between the networks of 15 May and 1 June 2021 is significant for all orbits. On the other hand, we find that the difference in the network starting on 1 October 2021 is driven primarily by changes in the distributions of cliques, through orbits 3 and 14, but we also smaller changes in orbits 0, 10 and 13.

We conclude this section with a final remark about the differences between the vaccine and COVID-19 datasets, a comparison that we show in Figure~\ref{fig:joined_dists}. We find that there is a clear distinction between the two sets of networks in the way they are organized, which indicates distinct user behaviours in each forum. This distinction is more tenuous when we consider the first network from the COVID-19 dataset, which is more dissimilar to the other networks in the COVID-19 dataset than to the networks in the vaccine dataset. This first network has an average distance to other COVID-19 networks of 0.18 and to vaccine networks of 0.21; the other COVID-19 networks have an average distance of 0.06 to COVID-19 networks and 0.20 to vaccine networks.

By analyzing the differences between the vaccine and COVID-19 datasets for each individual orbit (Figure~\ref{fig:joined_dists_allorbs}), we find that in 6 of the 14 orbits the COVID-19 dataset is more similar to the cluster of similar networks in the vaccine dataset before May 2015 than this cluster is similar to the networks of the vaccine dataset after May 2015. However, the opposite only happens in orbit 9. This highlights how fundamentally different user behaviour is before and after May 2015 in the vaccination forum, to the point where networks whose size differs by more than an order of magnitude are organized in a more similar way than networks of comparable size.

\subsection{Sentiment Analysis}

We combine three metrics to summarize sentiment in each time window:
\begin{itemize}
    \item Number of threads: the most direct metric from our labeling process, we measure the proportion of threads (i.e. original posts) labeled with each sentiment.
    \item Number of posts in thread: this allows us to gauge how popular each type of thread is, identifying cases where, for example, there are many threads with a particular sentiment label but each thread has very few posts; or where there are only a few threads with a sentiment label but each thread receives a large number of replies.
    \item Number of users: we measure which types of sentiment attract more users and how often users post in threads of different sentiment. This metric is useful for understanding how much information flows between threads of different sentiments.
\end{itemize}

For each of these metrics, we measure how the proportion of positive to negative, negative to positive and neutral to non-neutral sentiments changes over time. For example, in a given time window, we calculate the number of positive threads divided by the number of negative threads, the number of negative threads divided by the number of positive threads and the number of neutral threads divided by the number of positive or negative threads. We calculate these proportions for number of threads, number of posts and number of users, measuring how they vary over time.

\subsubsection{Vaccination forum}

\begin{figure}
    \centering
    \begin{subfigure}{\textwidth}
    \centering
    \includegraphics[width=0.65\textwidth]{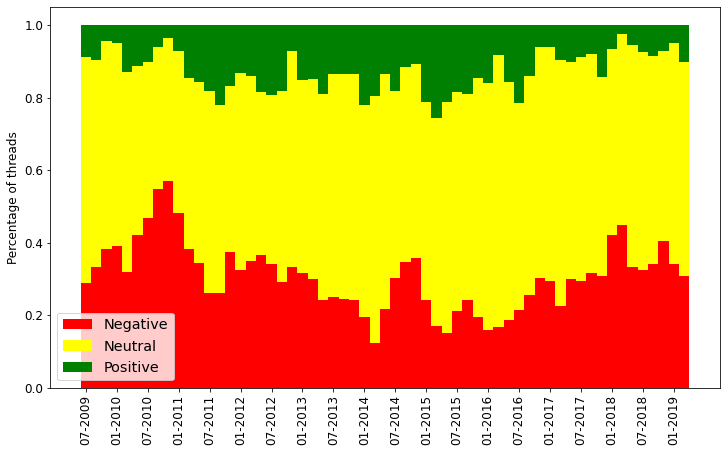}
    \caption{Proportion of threads}
    \label{fig:vaccine_thread_bars}
    \end{subfigure}
    \begin{subfigure}{\textwidth}
    \centering
    \includegraphics[width=0.65\textwidth]{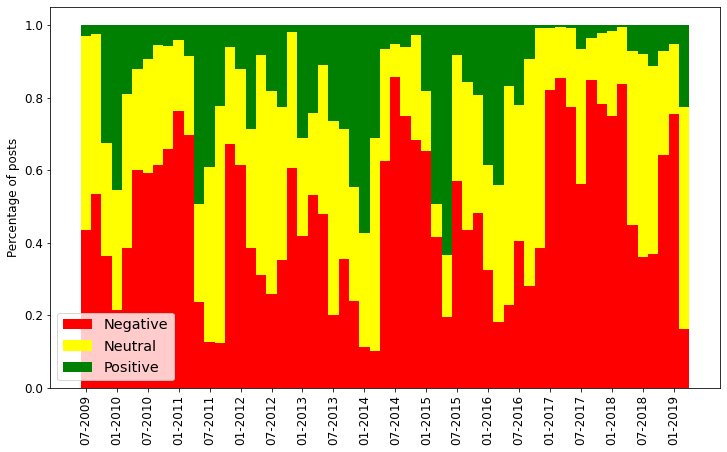}
    \caption{Proportion of posts}
    \label{fig:vaccine_posts_bars}
    \end{subfigure}
    \begin{subfigure}{\textwidth}
    \centering
    \includegraphics[width=0.65\textwidth]{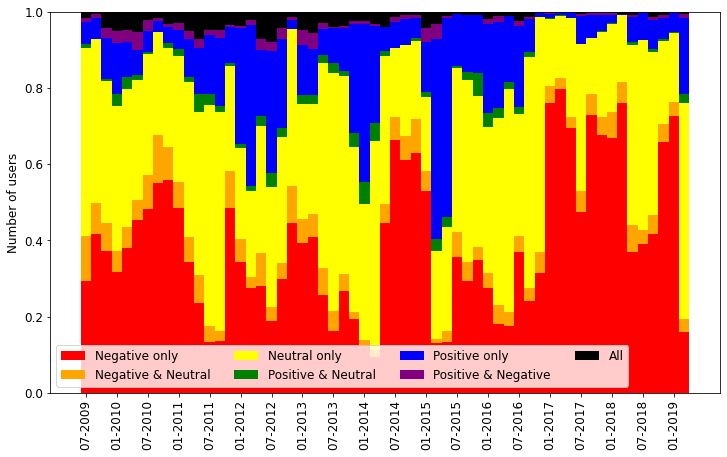}
    \caption{Proportion of users}
    \label{fig:vaccine_users_bars}
    \end{subfigure}
    \caption{Proportion of threads of each sentiment label, posts in threads of each sentiment label and users in threads of all combinations of sentiment label, for each data slice in the vaccination dataset.}
    \label{fig:vaccine_bars}
\end{figure}

Starting with the vaccination subforum, we show how the proportion of threads of each sentiment  (Figure~\ref{fig:vaccine_thread_bars}), the proportion of posts in threads of each sentiment  (Figure~\ref{fig:vaccine_posts_bars}) and the proportion of users in threads of all combinations of sentiment  (Figure~\ref{fig:vaccine_users_bars}), changes over time.


We find that the most common sentiment when opening a new discussion thread is neutral, with 58\% of threads labeled this way. One of the primary uses of the Mumsnet forum is for parents to seek advice from other parents, therefore many of the threads marked as neutral may be questions that do not incite discussion beyond answering the question. Evidence that supports this is that neutral threads tend to have shorter active times, with 70\% lasting less than a week, compared with 56\% and 52\% for positive and negative, respectively. The second most common sentiment is negative, with 29\% of discussions starting with a viewpoint against vaccination. The proportion of negative threads is particularly high before March 2013 and after November 2017,  with 37\% and 36\% of  threads being negative during these time periods respectively. The least common type of thread is positive (13\% of all threads), which may be because those who are content with the status quo (high rates of vaccination) are less likely to post than those who wish it to change.

The largest number of posts were written as replies to negative threads (44\%), compared with 36\% of replies in response to 
neutral threads and 20\% in response to positive threads. In spite of this, positive and negative threads receive on average a similar number of replies, 44 and 42 respectively, much higher than the average 17 replies per neutral thread. Figure~\ref{fig:vaccine_posts_bars} highlights how there are time periods dominated by discussions of each sentiment, for example from September 2017 to March 2018 78\% of posts were replies in negative threads but from May 2015 to May 2016 this proportion decreases to 28\%.

Figure~\ref{fig:vaccine_users_bars} shows that the majority of users -- 89\% -- post in only one type of thread. The biggest crossover between types of threads is neutral and negative, accruing an average of 6\% of users over time, against 1\% that post in both positive and negative, 2\% in positive and neutral and 2\% in threads of all sentiments. We find that after July 2015, the proportion of users that post in threads of all sentiments or in positive and negative threads decreases significantly; prior to this date these percentages are 3.5\% and 1.8\% respectively, decreasing to 0.6\% and 0.5\% after July 2015. 

These numbers are biased by the majority of users who post in only one thread. Across all data slices, on average only 15\% of users post in multiple threads but these users are responsible for 47\% of posts. When we consider only these users that post on at least two threads, the majority of them (69\%) post in threads of different sentiments, with 1\% posting only in positive threads, 19\% only in negative and 11\% only in neutral. As before, the most common combination is posting in negative and neutral threads, with 38\% of users doing so, followed by posting in threads of all sentiments (14\% of users).

To summarise, even though Figure~\ref{fig:vaccine_users_bars} shows that only a small proportion of users post in threads of different sentiments, when restricting the set of users to those that post in at least two threads, they are more likely to post in thread of different sentiments. As these users are responsible for nearly half of the posts on the board, it is reasonable to assume that there is information flowing between threads of different sentiments.

We are interested in knowing if the shifts we observe in sentiment correlate with differences in network structure. Thus, we need a systematic and quantifiable way to detect whether sentiment in one time window is significantly different from that in other time windows. To achieve this, for each of the metrics above, we calculate the proportion that is positive to negative, negative to positive and neutral to non-neutral, and then calculate a Z-score (as standard, through subtracting the mean and dividing by the standard deviation) for each proportion. 

When the Z-scored proportion takes a value above 1 or below -1 for multiple metrics related to the same sentiment within a time window, we consider that dimension to contain a significant change in sentiment. We choose the value of 1 (corresponding to two-sided statistical significance at the 32\% level for each metric) rather than the more common z-score absolute value of 2 (corresponding to two-sided statistical significance at the 5\% level for each metric) to avoid ignoring excessive numbers of changes due to the use of multiple metrics. To give an example of how our procedure works, an increase both in the proportion of neutral threads and posts in neutral threads is considered significant, but if the increase was only observed in the number of neutral threads, then that time window would not be considered significant. Figures~\ref{fig:vaccine_thread_sentiment},~\ref{fig:vaccine_posts_sentiment} and~\ref{fig:vaccine_users_sentiment} show this Z-score over time for threads, posts and users, respectively. These figures also contain all the network differences we mention in the previous section, marked as a black dotted line on index $i$ when the change occurs from index $i$ to $i+1$ and a gray dotted line when the change occurs from index $i$ to $i+2$. In the cases when both black and gray dotted lines ought to be present, we give precedence to the difference between consecutive time windows and show only the black dotted line.

According to this definition of significant change in sentiment, we observe a total of 20 (out of 59) time windows with a significant change in sentiment, compared to 16 differences in network structure.

We separate the differences in network structure into 6 groups:
\begin{itemize}
    \item The difference between the network starting in March 2012 and the network starting in May 2012 is associated with a decrease in the proportion of neutral threads (Figure~\ref{fig:vaccine_thread_neutral}) and an increase in the proportion of posts in neutral threads (Figure~\ref{fig:vaccine_posts_neutral}). There is also a significant increase in the proportion of users that post in multiple threads of different sentiments during this time (Figure~\ref{fig:vaccine_users_mult_sent}).
    \item The differences between the networks of May 2014 to September 2014 and of September 2014 to January 2015 do not appear to match with significant differences in sentiment during this period.
    \item The changes in network structure in March 2015 and May 2015 are associated with increases in the proportion of positive threads (Figure~\ref{fig:vaccine_thread_pos_neg}), posts in positive threads (Figure~\ref{fig:vaccine_posts_pos_neg}) and users posting only in positive threads (Figure~\ref{fig:vaccine_users_pos_neg}).
    \item The differences in March and May 2016 are correlated with increases in the proportion of neutral threads (Figure~\ref{fig:vaccine_thread_neutral}), posts in neutral threads (Figure~\ref{fig:vaccine_posts_neutral}) and users posting only in in neutral threads (Figure~\ref{fig:vaccine_users_neutral}).
    \item Between November 2016 and July 2017, we record differences of network structure in 5 consecutive time windows and during this time we observe an increase in the proportion of neutral threads (Figure~\ref{fig:vaccine_thread_neutral}), an increase in the proportion of posts in negative threads (Figure~\ref{fig:vaccine_posts_neg_pos}), an increase in the proportion of users posting only in negative threads (Figure~\ref{fig:vaccine_users_neg_pos}) and a decrease in the proportion of users posting in multiple threads of different sentiments (Figure~\ref{fig:vaccine_users_mult_sent}). There is also a significant shift from the time slice starting in November 2016 to January 2017 in the number of posts and users in neutral threads (Figures~\ref{fig:vaccine_posts_neutral} and~\ref{fig:vaccine_users_neutral}), from a proportion significantly higher than the average to significantly lower.
    \item The group of network differences between January 2018 and July 2018 is related to an increase in the proportion of negative threads (Figure~\ref{fig:vaccine_thread_neg_pos}), posts in negative threads (Figure~\ref{fig:vaccine_posts_neg_pos}) and users posting only in negative threads (Figure~\ref{fig:vaccine_users_neg_pos}).
\end{itemize}

We also find periods of time where there is change in the sentiment that is not reflected in differences in the network structure. In particular, we observe a high proportion of positive threads (Figure~\ref{fig:vaccine_thread_pos_neg}), posts in positive threads (Figure~\ref{fig:vaccine_posts_pos_neg}) and users posting only in positive threads (Figure~\ref{fig:vaccine_users_pos_neg}) in January and March 2014. The final time period that fits our definition of sentiment change but does not align with differences in network structure is between July and September 2011, where we detect a higher  than usual number of posts in positive threads (Figure~\ref{fig:vaccine_posts_pos_neg}), posts in neutral threads (Figure~\ref{fig:vaccine_posts_neutral}) and users in neutral threads (Figure~\ref{fig:vaccine_users_neutral}).

\subsubsection{COVID-19 forum}

\begin{figure}
    \centering
    \begin{subfigure}{\textwidth}
    \centering
    \includegraphics[width=0.65\textwidth]{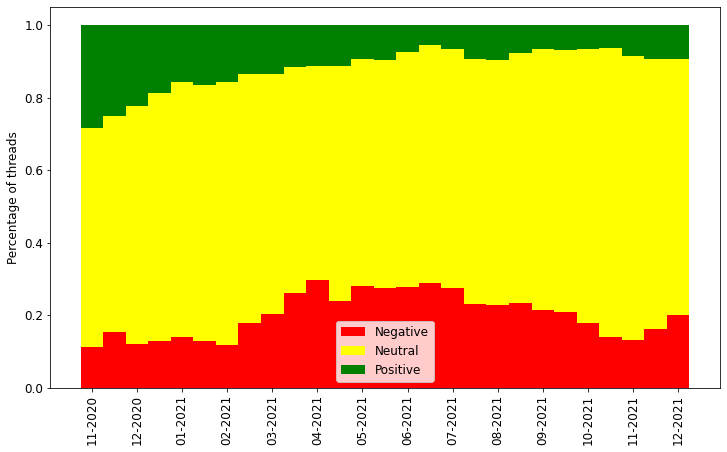}
    \caption{Proportion of threads}
    \label{fig:covid_thread_bars}
    \end{subfigure}
    \begin{subfigure}{\textwidth}
    \centering
    \includegraphics[width=0.65\textwidth]{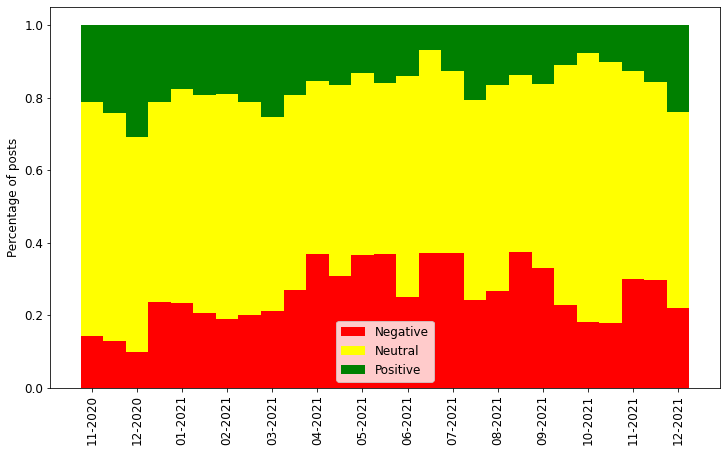}
    \caption{Proportion of posts}
    \label{fig:covid_posts_bars}
    \end{subfigure}
    \begin{subfigure}{\textwidth}
    \centering
    \includegraphics[width=0.65\textwidth]{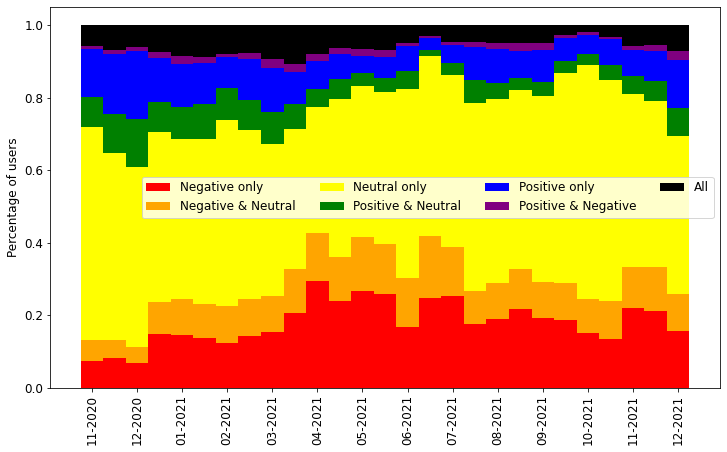}
    \caption{Proportion of users}
    \label{fig:covid_users_bars}
    \end{subfigure}
    \caption{Proportion of threads of each sentiment label, posts in threads of each sentiment label and users in threads of all combinations of sentiment label, for each data slice in the COVID-19 dataset.}
    \label{fig:covid_bars}
\end{figure}

We examine how the proportion of threads of each sentiment (Figure~\ref{fig:covid_thread_bars}), proportion of posts in threads of each sentiment (Figure~\ref{fig:covid_posts_bars}) and proportion of users in threads of all combinations of sentiment (Figure~\ref{fig:covid_users_bars}) change over time.

As with the vaccination forum, the most common sentiment when starting a thread about vaccination during COVID-19 is neutral (69\% of threads). A major difference to the vaccination forum is that threads with neutral sentiment are the most common in all data slices, whereas in the vaccination dataset there were time windows with more negative threads than neutral. The reason for this prevalence of neutrality may be that many threads about vaccination during the pandemic are not discussing the actual vaccine but instead issues surrounding the vaccination campaign, such as logistics or rules for vaccinated people, so the original post of these threads does not contain any judgement of value towards the vaccine, which we marked as neutral sentiment.

As before, the second most common sentiment is negative (20\% of threads). Negative sentiment is particularly high between March and July 2021, which coincided with the Astra Zeneca blood clot issue and users complaining about side effects of vaccines. The least common sentiment is positive (11\% of threads). However, the first four time windows with a start date in 2020 have a much higher proportion of positive threads compared with the remainder of the dataset, with an average of 23\% positive threads over this time period.

Although they are few in number, positive threads attract the highest number of replies, averaging 70 replies per time window, against 60 in negative threads and 39 in neutral. Despite attracting more replies, the low number of threads means that posts in positive threads are nonetheless the minority of all posts, averaging 17\% of all posts over time, as illustrated in Figure~\ref{fig:covid_posts_bars}. The popularity of individual positive threads can be explained by ``positivity" threads, created to be a safe space where users can feel good about the progress of the vaccination campaign.

The prevalence of neutral threads leads to posts in these constituting the majority of posts in all time windows, with an average of 57\% over time. Posts in negative threads make up the remaining 26\% and are the ones with the greatest variance in proportion over time.

Neutral threads contain the greatest proportion of users, with 48\% of users posting only in neutral threads, 11\% in negative and neutral, 6\% in positive and neutral and 6\% in threads of all sentiments. The remaining 29\% of users are split between 18\% that only post in threads of negative sentiment, 9\% in threads of positive sentiment and 2\% in threads of either positive or neutral sentiment.

Figure~\ref{fig:covid_users_bars} exposes another difference in behaviour between users discussing vaccination in the COVID-19 subforum and in the defunct general vaccination subforum. Whereas in the latter we found that 89\% of users restricted themselves to only one of positive, neutral or negative threads, in the COVID-19 dataset this number decreases to 75\%. This is likely because the proportion of users that post in multiple threads is higher in the COVID-19 dataset (35\% against 15\% in the vaccination dataset) and they make up a larger proportion of the total number of posts (75\% against 47\%). In spite of this difference, when considering the users that post in multiple threads, the same percentage (69\%) of these post in threads of different sentiment in both datasets.

As with the vaccination dataset, we want to understand whether changes in sentiment are associated with differences in network structure. We previously saw that the majority of networks in the COVID-19 dataset are very similar, which fits with the results we have just discussed, as the prevailing sentiment throughout the 13 months of data is of neutrality. Nonetheless, we observe both differences in network structure and in the proportion of negative, positive and neutral threads and of posts and users therein. Using the same methodology as before, we show how the proportion of negative to positive, positive to negative and neutral to positive plus negative changes over time in regard to the number of threads (in Figure~\ref{fig:covid_thread_sentiment}), posts (Figure~\ref{fig:covid_posts_sentiment}) and users (Figure~\ref{fig:covid_users_sentiment}).

We categorise the network differences in 4 groups:
\begin{itemize}
    \item The first two networks are significantly different from the remaining networks in the dataset and this difference is connected to a higher representation of positive sentiment  (Figures~\ref{fig:covid_thread_pos_neg},~\ref{fig:covid_posts_pos_neg} and~\ref{fig:covid_users_pos_neg}) and a lower representation of negative sentiment (Figures~\ref{fig:covid_thread_neg_pos},~\ref{fig:covid_posts_neg_pos} and~\ref{fig:covid_users_neg_pos}) in all three metrics.
    \item The difference between the networks starting on 15 March and 1 April 2021 coincides with a decrease in the proportion of neutral threads (Figure~\ref{fig:covid_thread_neutral}), posts in neutral threads (Figure~\ref{fig:covid_posts_neutral}) and users in neutral threads (Figure~\ref{fig:covid_users_neutral}). The time window starting on 15 March is also the time with the highest proportion of users posting in multiple threads with different sentiments (Figure~\ref{fig:covid_users_mult_sent}).
    \item The difference between the networks of 1 May and 1 June and between 15 May and 1 June 2021 are both above the $80th$ percentile of network distances, but we find little evidence of sentiment change in these time periods, despite the slight underrepresentation of posts and users in neutral threads (Figures~\ref{fig:covid_posts_neutral} and~\ref{fig:covid_users_neutral}) during the time window starting on 15 May. We also observe network differences between the networks of 15 May and 15 June and between 1 June and 1 July 2021, which can be related to the high proportion of negative sentiment in the data slices of 15 June and 1 July.
    \item The final networks of the COVID-19 dataset, after 1 September 2021, show a low degree of similarity, either in consecutive networks (networks starting on 15 September, 15 October and 15 November 2021) or in networks two jumps apart (networks starting 1 September, 1 October and 1 November 2021). These differences are likely to be related to a high proportion of neutral sentiment during this time, in particular the two time windows of 1 and 15 October 2021, or a low proportion of users posting in multiple threads with different sentiments.
\end{itemize}
Although we find some overlap between differences in network structure and changes in the sentiment of the forum, these do not correspond to a one to one match. For instance, the highest proportion of posts and users in positive threads occurs in 1 December 2020, but there the network structure at that point is similar to the structure of the networks that succeed it. Conversely, some network differences do not seem to be connected with changes in sentiment, such as the difference between 15 May and 1 June 2021 that we mentioned previously or the difference between 15 November and 1 December 2021.

Due to these misalignments between differences in network structure and sentiment variation, we decided to label all posts from a subset of threads from the COVID-19 forum. The threads were picked randomly, with equal proportion of positive, negative and neutral threads. Some time windows were underrepresented in the initial sample, so we re-sampled more threads from each time window that was underrepresented until there was at least 1\% of threads and posts from each time window in the sample. This led to a total of 129 threads (about 3.0\% of the number of threads) which contained 6898 posts (about 3.1\% of the number of posts).

The posts in this dataset were labeled using the same rules as the ones we used to label the original post of each thread. Upon labeling these posts, we calculate the proportion of posts of each sentiment in threads of each sentiment. Table~\ref{tab:covid_posts_prop} shows this average proportion for each combination in the threads that we labeled. The average shown is the \emph{macro} average, i.e., the average proportion of each thread; results do not change substantially when using the \emph{micro} average instead.

\begin{table}
\centering
  \renewcommand{\arraystretch}{1.5}
  \begin{tabular}{>{\centering\arraybackslash}m{15pt}>{\centering\arraybackslash}m{55pt}>{\centering\arraybackslash}m{55pt}>{\centering\arraybackslash} m{55pt}>{\centering\arraybackslash} m{55pt}}
    & & \multicolumn{3}{c}{\large Threads} \\
    & & Positive & Neutral & Negative\\
   \multirow{3}{*}{\rotatebox[origin=c]{90}{\large Posts}} &  
    Positive & \multicolumn{1}{c}{51.7\%} & 19.0\% & 22.5\%\\
    & Neutral & \multicolumn{1}{c}{42.0\%} & 76.9\% & 47.3\% \\
    & Negative & \multicolumn{1}{c}{6.3\%} & 4.1\% & 30.2\%\\
  \end{tabular}
  \caption{Proportion of posts of positive, neutral or negative sentiment in threads of positive, neutral or negative sentiment. The table is read as: posts of sentiment A make up X\% of the posts in threads of sentiment B. For example, posts of positive sentiment make up 22.5\% of the posts in threads of negative sentiment or posts of negative sentiment make up 6.3\% of the posts in threads of positive sentiment.}
  \label{tab:covid_posts_prop}
\end{table}

The results from Table~\ref{tab:covid_posts_prop} show that negative threads have the widest range of sentiments, as users with opinions favourable towards vaccination often challenge negative views. On the other hand, despite their smaller frequency, positive threads are havens for users who wish to shield themselves from negativity in the forum, as negative threads are more frequent than positive ones. In positive threads, we find that the majority of posts are also positive towards vaccination, likely due to ``positivity" threads that celebrate events like new vaccines being approved or milestones in the number of doses administered. Of all sentiments, neutral threads have the smallest proportion of negative posts, perhaps because they are either question and answer threads that do not evolve into lengthy discussions and thus are not attractive to users who want to spread an anti-vaccination agenda; or they turn into discussions about issues other than vaccination, for example whether restrictions should be lifted.

\begin{figure}
    \centering
    \includegraphics[width=0.85\textwidth]{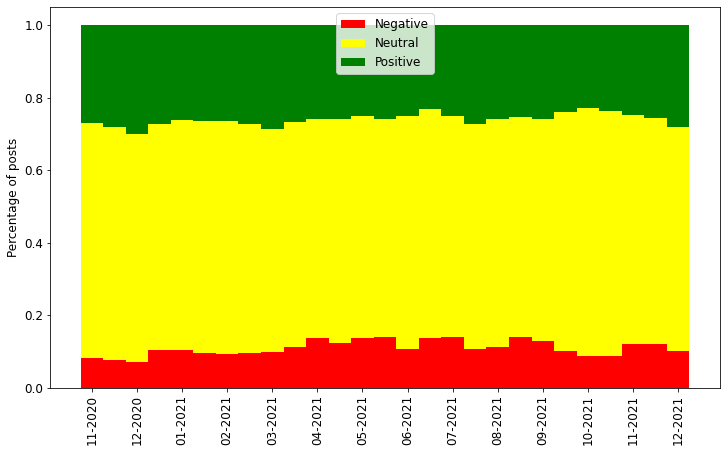}
    \caption{Percentage of posts of each sentiment label, inferred from the proportion of posts of each sentiment in threads of each sentiment in the labeled sample of the COVID-19 dataset, results shown for each data slice.}
    \label{fig:covid_posts_bars_hand}
\end{figure}

Assuming that the proportions from Table~\ref{tab:covid_posts_prop} are representative of the whole population of threads in our dataset, we can infer the percentage of posts of each sentiment over time, shown in Figure~\ref{fig:covid_posts_bars_hand}.

Even though there are an overall higher number of posts in negative threads than positive threads, a higher proportion of posts are expressing a positive sentiment than a negative one. We find that 26\% of posts convey a positive sentiment, which is substantially greater than the 17\% of posts written as replies to positive threads. Conversely, the proportion of posts in negative threads was an overestimation of the total percentage of negative posts in the dataset, which we calculate to be an average of 11\% across all time slices. The proportion of posts in neutral threads was also an underestimation for the overall percentage of neutral posts, changing from 57\% to 63\%.

Employing the same methodology as before of Z-scoring the proportion of positive to negative, negative to positive and neutral to positive plus negative posts, we are able to identify a possible reason for the network structure difference between the network of 1 June 2021 to its neighbours. As can be seen in Figure~\ref{fig:covid_posts_hand_sentiment}, in the data slices of 1 May, 15 May, 15 June and 1 July 2021 there is an underrepresentation of positive posts and an overrepresentation of negative ones, which is not the case for the time window starting on 1 June 2021. The same effect happens for the number of users in negative threads (Figure~\ref{fig:covid_users_neg_pos}), which are overrepresented in the four closest time windows to 1 June 2021, but not in the time window of 1 June 2021.

\begin{figure}
    \centering
    \includegraphics[width=0.85\textwidth]{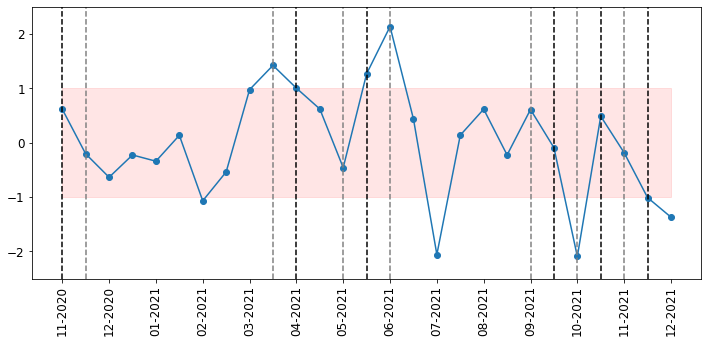}
    \caption{Average discordance index per thread in the sampled subset of the COVID-19 dataset over time, calculated as the average discordance using windows with 2 to 5 posts.}
    \label{fig:covid_discordance}
\end{figure}

Finally, we explore the level of disagreement in each thread of the sampled subset of the COVID-19 dataset over time, through the discordance index metric we propose. Figure~\ref{fig:covid_discordance} shows the average discordance per thread over time, with the discordance for each thread calculated as the average discordance using windows of 2 to 5 posts.

Results using the discordance index indicate that the network difference we observe in 1 June 2021 may be connected with a period of a high degree of disagreement within the forum. Another period of time that stands out is the particularly low level of disagreement in the time window starting on 1 October 2021, which we also previous highlight as it coincides with a significant difference in network structure. On the other hand, the discordance index suggests that during the time window of 1 July 2021 there is a very low level of disagreement that does not coincide with a difference in network structure.

\begin{figure}
    \centering
    \includegraphics[width=0.85\textwidth]{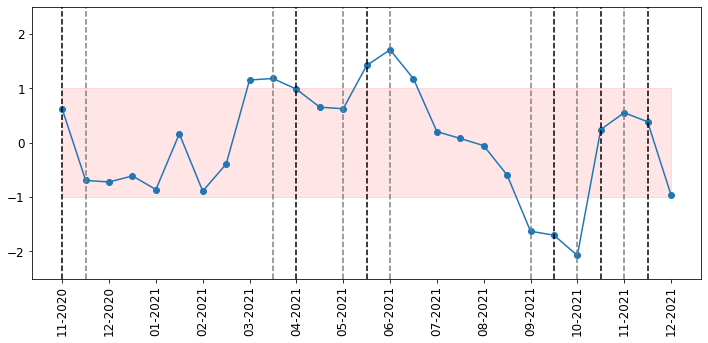}
    \caption{Average discordance index per thread in the COVID-19 dataset over time, calculated as the average discordance using windows with 2 to 5 posts. Values are inferred from the sample of sentiment labeled posts.}
    \label{fig:covid_discordance_infer}
\end{figure}

Assuming our sample is representative of the whole dataset, we can calculate the average discordance value in threads of each sentiment type and use it to infer the average discordance index for the whole dataset, which we show in Figure~\ref{fig:covid_discordance_infer}.

This methodology still emphasizes 1 June 2021 as a time of particularly high level of disagreement in the forum. The results also uncover a possible explanation for the network structure changes after September 2021, as the time windows starting on 1 September, 15 September and 1 October 2021 have the lowest levels of disagreement of all time frames and they are followed by a rapid increase in the discordance index in 15 October 2021.

\section{Discussion}

\subsection{Summary of results}

We have shown how user interactions with the Mumsnet forum can be represented through complex networks. By splitting these interactions according to their date, we are able to obtain snapshots of these interactions over a fixed period of time. We employed the technique of network comparison to observe differences between each snapshot of interactions, by measuring how the structure of the network obtained from these interactions changes over time.

We found that differences in network structure often occur simultaneously to real world events that affect discussions in Mumsnet. For instance, the three major differences in the network structure of the COVID-19 forum coincide with the approval of the Pfizer and Astra Zeneca vaccines, the emergence of the Delta variant and the start of the vaccine booster program. Some of the differences in network structure we identify can be traced to changes in the distribution of specific orbits, which form a basis for the interpretation of the reasons for the differences in structure. An example of a change in the distributions of a subset of orbits that leads to significant differences in the structure of the whole network are the distributions of cliques in the vaccination subforum; these orbits are the biggest contributors for the separation of the networks in this dataset into two broad clusters, before and after May 2015.

By performing sentiment labeling and analysis on these datasets, we were also able to uncover a link between differences in the network structure and changes in the sentiment toward vaccination in the forum. Out of 10 groups of network changes, only in one do we not find any connection to change in sentiment. Conversely, there are only two changes in sentiment that are not reflected in differences of network structure, both of which occur in the vaccination dataset.

Through a sample of the COVID-19 dataset, we studied whether differences in network structure were a better indicator of the distribution of sentiment within the posts of a thread. The distribution of sentiment in a thread can be seen as a proxy for the amount of disagreement within that discussion; although posts of the same sentiment do not necessarily indicate agreement, posts of different sentiment reflect opposing views. By using this metric of disagreement, results show that the time with most disagreement in our sample coincides with the second largest difference in network structure.

Considering Mumsnet as a whole, our analysis shows clear distinctions between how users discuss vaccination in general and how they discuss it within the context of COVID-19. This distinction starts in the way that discussions progress in each thread, with the COVID-19 forum characterized by short-lived threads with a high volume of replies. This contrasts with more drawn out threads, some that even last multiple years, in the vaccination forum. There are also big differences between the two forums in terms of the sentiment towards vaccination. Because the data for the vaccination forum spans a much longer period of time, it is natural that it contains greater shifts in sentiment over time. However, there is a much higher prevalence of neutral sentiment in the COVID-19 forum, as neutrality is the dominant sentiment in all time windows; the same does not happen in the vaccination forum, where there are time periods dominated by negative sentiment. Another key difference between the two forums is the number of users posting in threads of different sentiments, which can be seen as a proxy for the amount of information flowing between threads of different types. The proportion of users posting in threads of all sentiments is significantly higher in the COVID-19 forum than in the vaccination forum, which can be interpreted as an indication that there is a wider variety of opinions in each thread of the COVID-19 forum.

Underlying all these distinctions between the two forums, we find significant differences between the network structure of their user interactions. Although the contrast in network structure is to be expected for the reasons above, we observe unexpected similarities in the distributions of certain orbits. These similarities highlight common patterns of user interactions in Mumsnet as a whole and further accentuate the difference between the networks before and after May 2015, in the vaccination forum.

\subsection{How can network structure help us understand sentiment?}

Sentiment labeling is a very laborious process that consumes a great deal of human resources and does not scale to large datasets, as an increase in the quantity of data requires a corresponding increase in the effort of labeling. Automatic methods have been developed but the vast majority focuses on Twitter data and are unsuitable for more nuanced arguments that do not fit the character limit of a tweet. Furthermore, automatic methods require representative samples for training and pre-labeled data for training is often inappropriate due to changes in the context in which the discussions took place.

On the other hand, the network comparison methodology we use to analyze user behaviour in vaccination discussions is an easy to use out-of-the-box tool that scales to large social networks. Through network comparison, we were able to identify periods of time with differences in user behaviour that were associated with changes in sentiment towards vaccination. In other words, our results indicate that shifts in opinion are also reflected in changes to the network structure. Our methodology naturally does not replace sentiment analysis, but instead it can be useful for researchers as a tool to identify shifts in sentiment that can help to target further analysis.

Another strength of our approach of network comparison is that we are able to inspect which graphlets and orbits are responsible for the differences in network structure. This knowledge can be used to compliment sentiment analysis to generate hypothesis about differences in user behaviour.

We now provide an example of this ability to generate hypotheses from differences in network structure identified by the network comparison methods. In Section~\ref{sec:covid_net_struct}, we highlighted that the difference in the network starting on 1 October 2021 was being primarily driven by changes in the distributions of cliques, either of size 3 or 4. This change does not stem from an over or under representation of these subgraphs, as motif analysis indicates that the subgraph ratio profile~\citep{milo2004superfamilies} is very similar for these two networks. Instead, in October 2021 we observe that more users participate in a greater number of cliques, compared to the previous time slices where we see a smoother distribution of participation in cliques over the range of values. Using this information, we can hypothesize that the network from 1 October 2021 is composed of a large set of users densely connected and this large set of users is sparsely connected to the remaining users, who form their own small clusters. 

Building on this hypothesis, we can hypothesize about user behaviour that led to this network topology. Recall from Section~\ref{sec:net_creation} that this network of interest is a sparsified one-mode projection on the user partition from a user-thread bipartite network. Therefore, to reach this network topology, the bipartite network needs to have a small number of nodes from the thread partition connected to a large set of nodes from the user partition. From a social network perspective, this is equivalent to saying that there is a small number of threads where the majority of vaccination discourse was concentrated. This can indicate a sort of preferential attachment~\citep{barabasi1999emergence} towards popular threads, with mechanics resembling a Chinese restaurant process~\citep{aldous1985exchangeability}.

\subsection{Limitations}

Although we are able to judge which graphlets are responsible for the changes in network structure, we were unable to determine whether there is a correspondence between changes in certain graphlets and the prevalence of a specific sentiment. We tried alternative techniques, such as motif analysis through the comparison against a null model or even simply inspecting the frequency of graphlets, but changes in network structure are often quite subtle; it is not the prevalence of each subgraph that leads to the differences in network structure, but rather how they are distributed over the network. This is what makes complex networks complex; characterizing their structure is often hard so it is important to develop heuristics that can help identify when the structure is different compared to other networks that are hypothesized to be similar. Nevertheless, the relationship between network structure and sentiment is still an open problem and we leave the application of other techniques to uncover this relationship for future work.

One of the biggest limitations of our analysis of the COVID-19 forum is the wide criteria we used to include threads in our dataset, as all threads that included any keyword related to vaccination were considered in the analysis. This lead to the inclusion of threads where the discussion was not focused on the vaccines themselves but instead related to the pandemic in general. This wide criteria was particularly evident when labeling the sampled posts, where we identified discussions about lockdowns and opening up the country after the vaccination campaign, about Brexit in the context of Astra Zeneca vaccine distribution and other pandemic specific topics like modeling, masking and testing. These discussions were labeled as neutral towards vaccination but there was a large body of arguments in regard to each of these issues, with polarizing opinions between users.

It was unclear how to deal with this issue. Even if on a thread level it was possible to detect threads whose discussion was not aimed at the vaccines in particular when we labeled each original post, we found that even in threads that started off as discussions about the vaccine, the natural flow of discussion trended towards other issues. This was especially common in threads with a large number of posts, where off-topic concerns kept being brought up, usually as a reflection of the pandemic situation at the time.

These side discussions present a confounding factor in connecting the structure of the networks with the sentiment of the forum. The structure of the network carries information about how the discussions are being carried out, regardless of their content, whereas all the side discussions were grouped under the umbrella sentiment of neutrality. This affects all facets of our analysis but in particular the discordance index, as we observed long sequences of neutral posts that hid a high level of disagreement.

Another issue that affects both the COVID-19 and vaccination forums is moderation by Mumsnet staff in an effort to remove blatant misinformation. This is reflected in threads being deleted or posts whose content was deleted, as well as replies that quoted those deleted posts. Whereas the former only affects the overall perception of how much negative sentiment is in the forum, as naturally only threads negative towards vaccination were deleted, the deleted posts are still encoded in the network structure and were marked as neutral sentiment. The reason for marking deleted posts as neutral instead of negative is that they do not necessarily express anti-vaccination views borne out of misinformation, but instead often they are unfruitful discussions between users that involve personal attacks and thus violate the rules of the forum.

\section{Conclusion}

This paper presents a novel approach to the analysis of vaccination discussions in online social networks. In particular, we study how these discussions are carried out in Mumsnet, a discussion board targeted towards parents. This medium is often overlooked by researchers in favour of more popular social networks like Facebook or Twitter, but studies have found that discussion boards such as Mumsnet have a greater effect on vaccination decisions than Twitter and are a reflection of public opinion.

By recording users' posting patterns in threads, we are able to construct networks of user relationships, which represent how users behave and interact among themselves in the forum, over fixed lengths of time. One way to measure how these relationships change over time is by employing network comparison through distributions of graphlets. The methods we use to compare these networks require little parametrization, making them usable as out-of-the-box tools that scale for social networks with millions of users and are interpretable as they show which functional structures are responsible for the differences in structure.

We validate the output of the network comparison measure against indicators of sentiment towards vaccination in the forum, after hand labeling the original post of each thread related to vaccination. Results show that there is an association between differences in the network structure of user interactions and changes in the sentiment of threads. Thus, network comparison can be used as a tool to reduce the effort of sentiment labeling, a labour intensive and often expensive process, by providing a high level indication of when sentiment is changing.

\backmatter



\section*{Declarations}

\subsection*{Funding}

MEPS is funded by Engineering and Physical Sciences Research Council Manchester Centre for Doctoral Training in Computer Science (grant number EP/I028099/1). TH was supported by the Royal Society (INF\textbackslash{}R2\textbackslash{}180067) the JUNIPER modelling consortium (MR/V038613/1) and the Engineering and Physical Sciences Research Council (EP/V027468/1).

\subsection*{Compliance with Ethical Standards}

The authors declare no competing interests.

The data used in this study was acquired with the permission of the intellectual property holder, Mumsnet. The study was approved by the University of Manchester computer science department panel of ethics, with reference 2020-8214-12903.

\subsection*{Data availability}

The datasets generated during and/or analysed during the current study are not publicly available due to being intellectual property of Mumsnet but are available from the corresponding author on reasonable request.

\subsection*{Author's contributions}

Conceptualization: MEPS, TH, CJ; Methodology: MEPS, RS, TH, CJ; Formal analysis and investigation: MEPS, RS; Writing - original draft preparation: MEPS; Writing - review and editing: MEPS, RS, TH, CJ; ; Supervision: TH, CJ.

\clearpage

\begin{appendices}

\section{Data granularity visualization}

\begin{figure}[h]
    \centering
    \includegraphics[width=\textwidth]{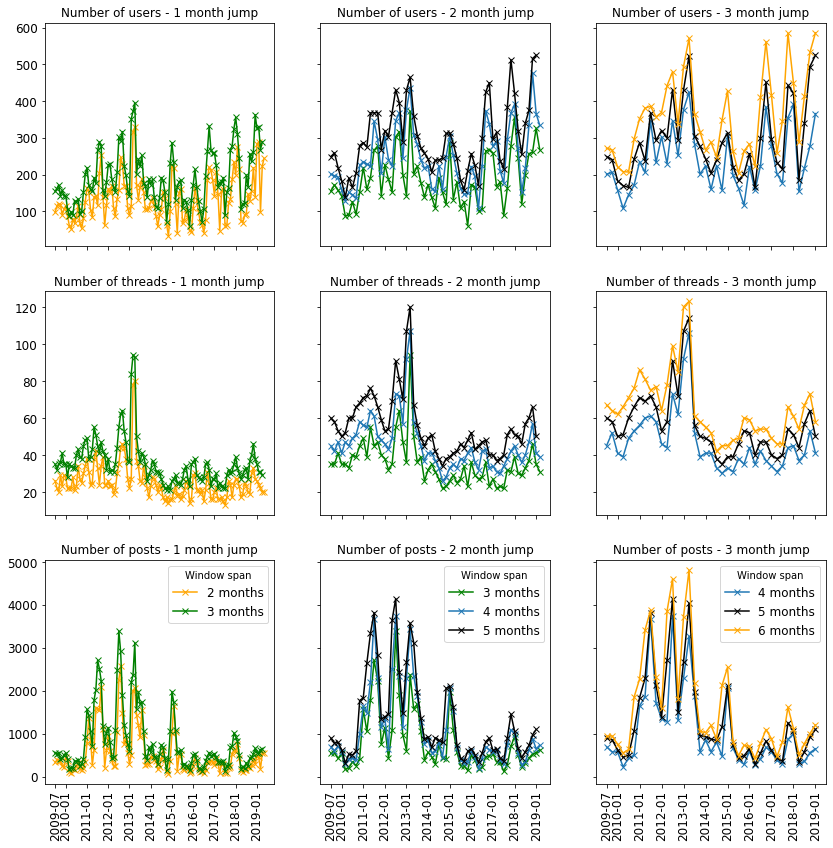}
    \caption{Number of users, threads and posts for different combinations of time jump and time span to define the time granularity in the dataset related to the vaccine forum.}
    \label{fig:vaccine_data_granularity}
\end{figure}

\begin{figure}
    \centering
    \includegraphics[width=\textwidth]{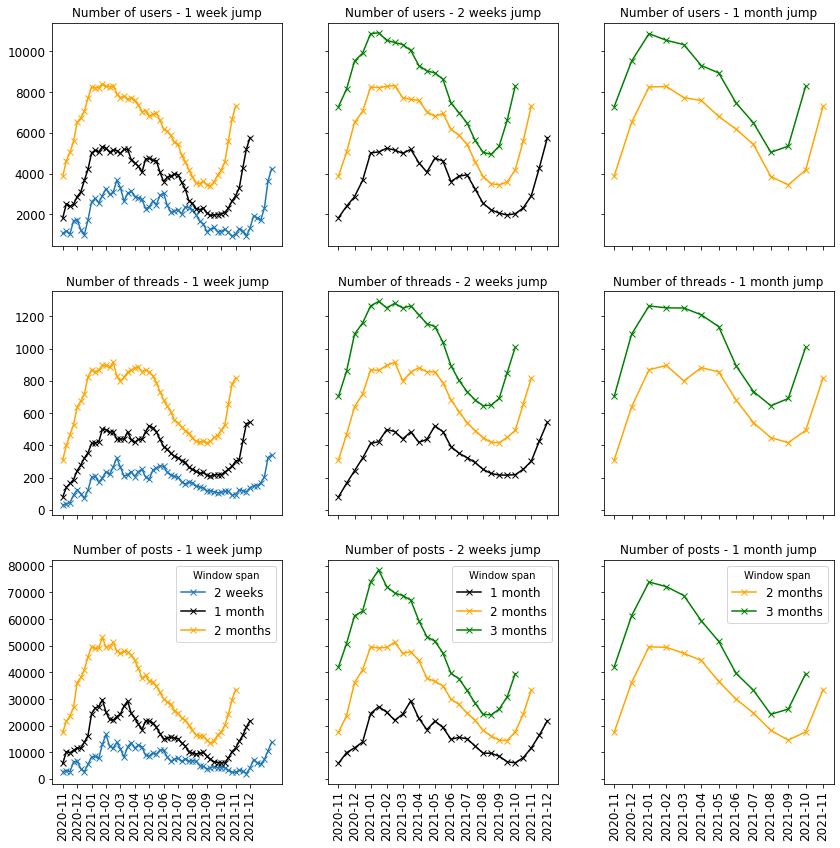}
    \caption{Number of users, threads and posts for different combinations of time jump and time span to define the time granularity in the dataset related to the COVID-19 forum.}
    \label{fig:covid_data_granularity}
\end{figure}

\clearpage

\section{Network distances with all individual orbits}\label{app:morefigs}

\begin{figure}
    \centering
    \includegraphics[width=\textwidth]{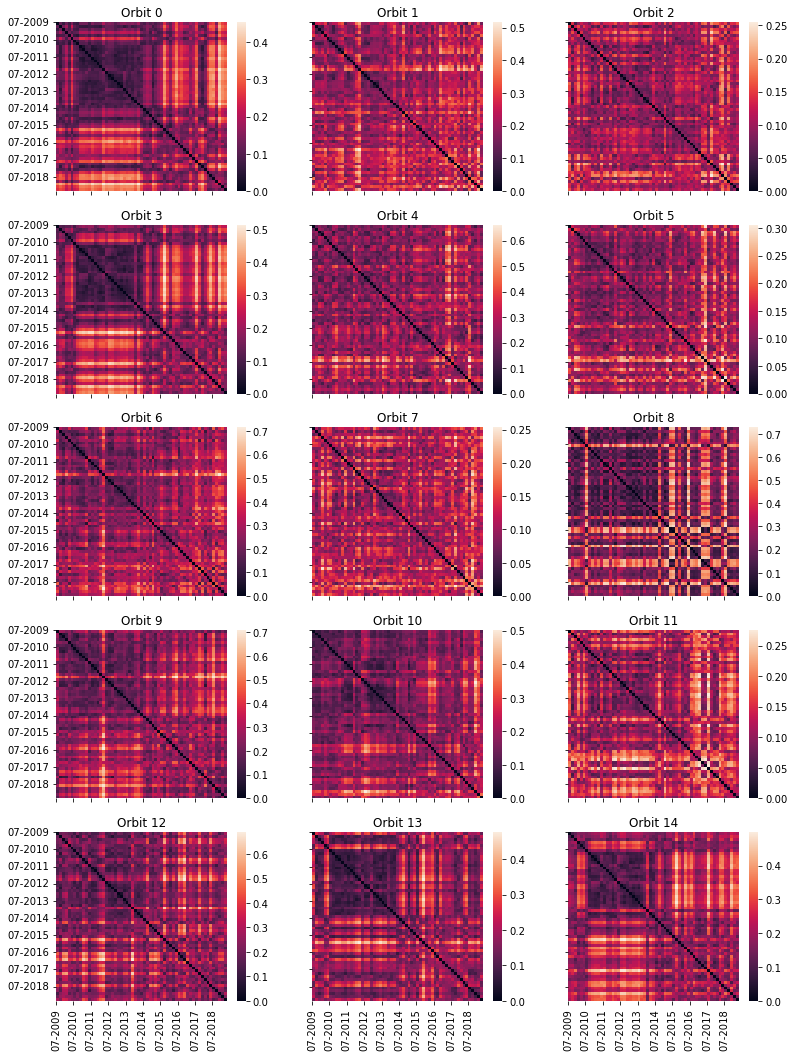}
    \caption{Heat map of network distances between each pair of networks in the Mumsnet vaccination dataset, for each orbit in graphlets of size up to 4. Distances are calculated using $PCA\_NetEmd$ \citep{silva2022comparing} with size 4 graphlets and 90\% explained variance.}
    \label{fig:vaccine_dists_allorbs}
\end{figure}

\begin{figure}
    \centering
    \includegraphics[width=\textwidth]{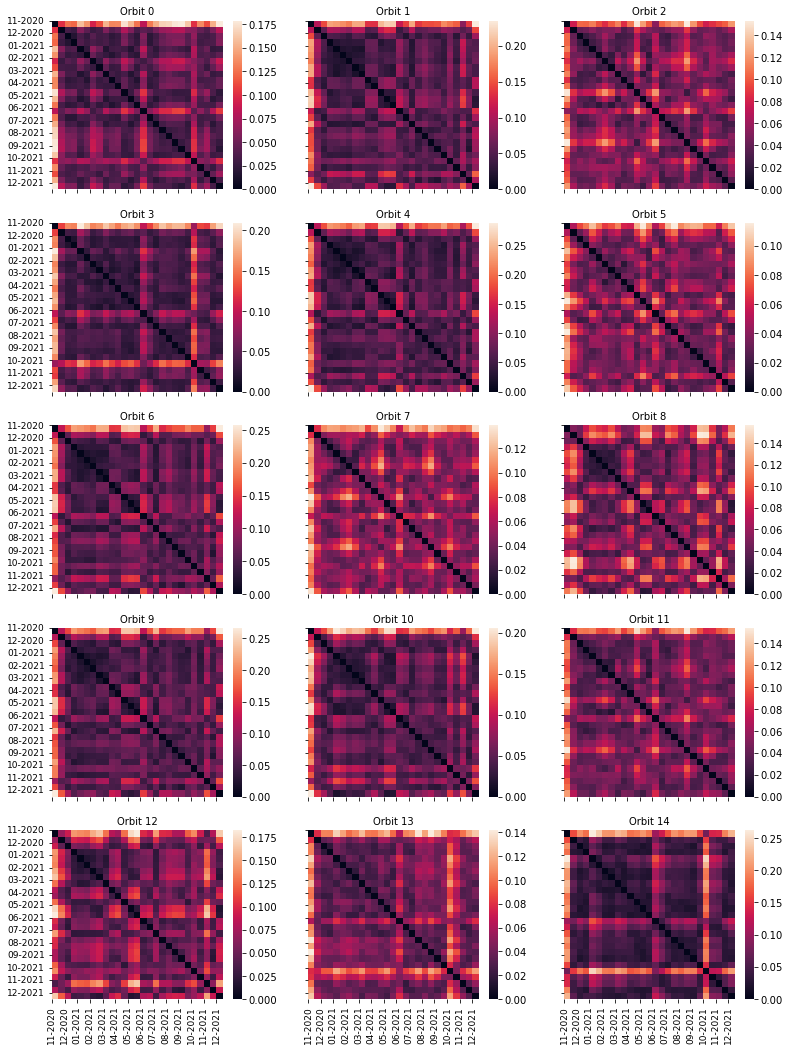}
    \caption{Heat map of network distances between each pair of networks in the Mumsnet COVID-19 dataset, for each orbit in graphlets of size up to 4. Distances are calculated using $PCA\_NetEmd$ \citep{silva2022comparing} with size 4 graphlets and 90\% explained variance.}
    \label{fig:covid_dists_allorbs}
\end{figure}

\begin{figure}
    \centering
    \includegraphics[width=\textwidth]{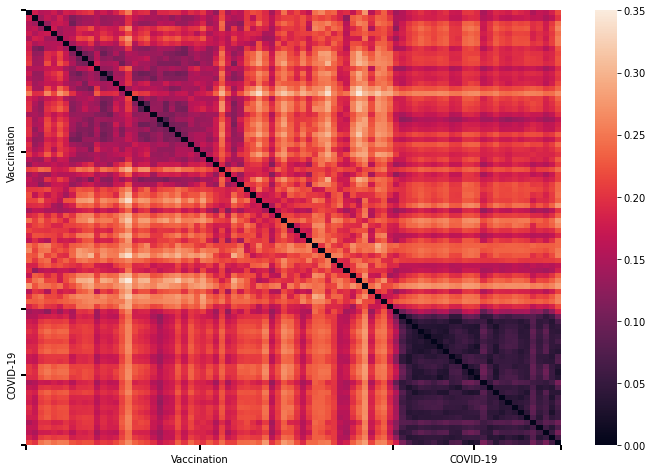}
    \caption{Heat map of network distances between all networks in both the Mumsnet vaccination and COVID-19 dataset. Distances are calculated using $PCA\_NetEmd$ \citep{silva2022comparing} with size 4 graphlets and 90\% explained variance.}
    \label{fig:joined_dists}
\end{figure}

\begin{figure}
    \centering
    \includegraphics[width=\textwidth]{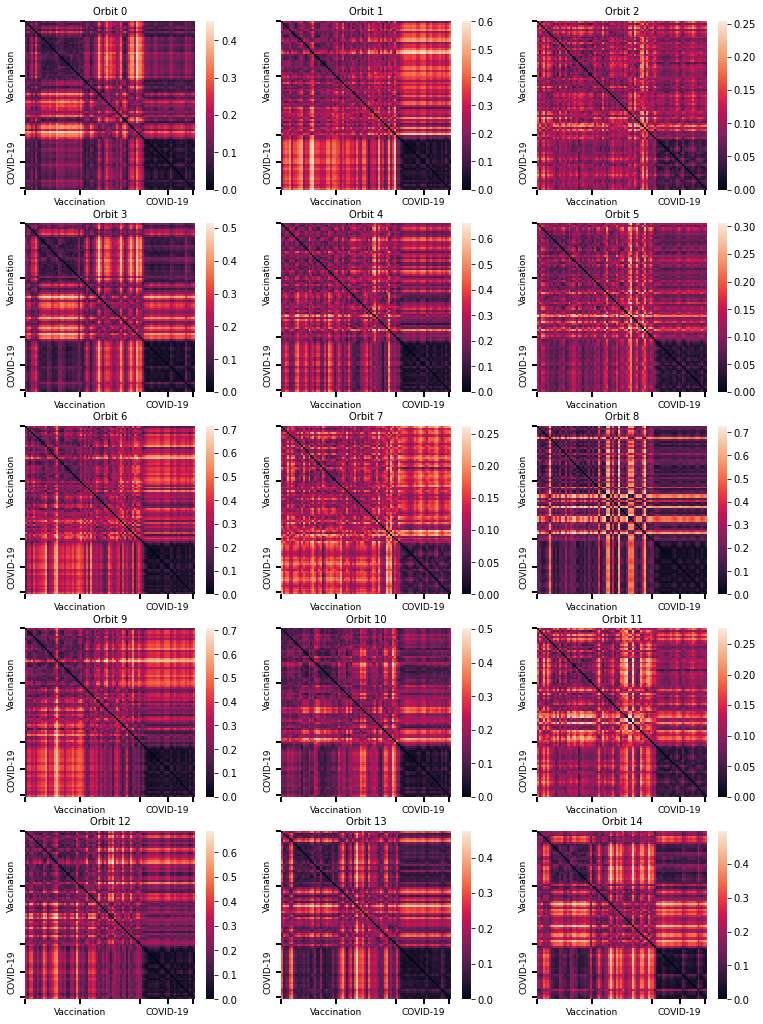}
    \caption{Heat map of network distances between all networks in both the Mumsnet vaccination and COVID-19 dataset, for each orbit in graphlets of size up to 4. Distances are calculated using $PCA\_NetEmd$ \citep{silva2022comparing} with size 4 graphlets and 90\% explained variance.}
    \label{fig:joined_dists_allorbs}
\end{figure}

\clearpage

\section{Sentiment analysis results}

\begin{figure}
    \centering
    \begin{subfigure}{\textwidth}
    \centering
    \includegraphics[width=0.9\textwidth]{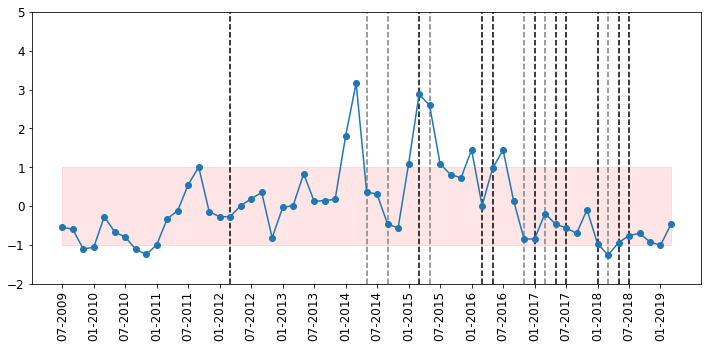}
        \caption{Proportion of positive to negative threads.}
        \label{fig:vaccine_thread_pos_neg}
    \end{subfigure}
    \begin{subfigure}{\textwidth}
    \centering
    \includegraphics[width=0.9\textwidth]{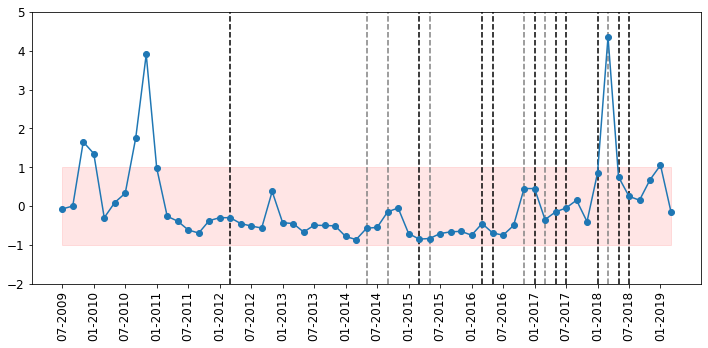}
        \caption{Proportion of negative to positive threads.}
        \label{fig:vaccine_thread_neg_pos}
    \end{subfigure}
    \begin{subfigure}{\textwidth}
    \centering
    \includegraphics[width=0.9\textwidth]{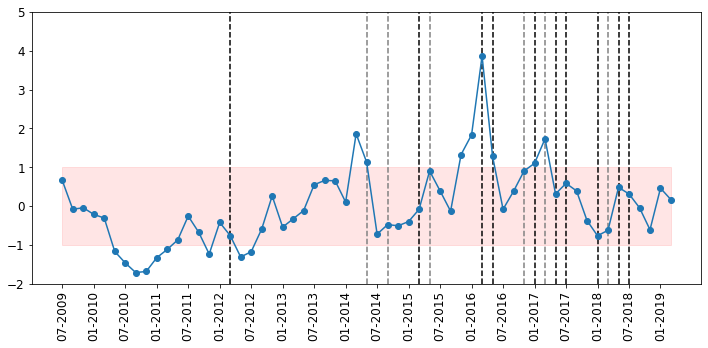}
        \caption{Proportion of neutral to positive and negative threads.}
        \label{fig:vaccine_thread_neutral}
    \end{subfigure}
    \caption{Vaccination subforum - threads of each sentiment.}
    \label{fig:vaccine_thread_sentiment}
\end{figure}

\begin{figure}
    \centering
    \begin{subfigure}{\textwidth}
    \centering
    \includegraphics[width=0.9\textwidth]{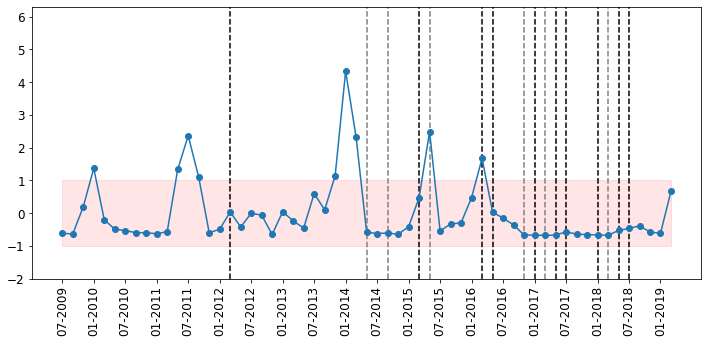}
        \caption{Proportion of posts in positive threads to posts in negative threads.}
        \label{fig:vaccine_posts_pos_neg}
    \end{subfigure}
    \begin{subfigure}{\textwidth}
    \centering
    \includegraphics[width=0.9\textwidth]{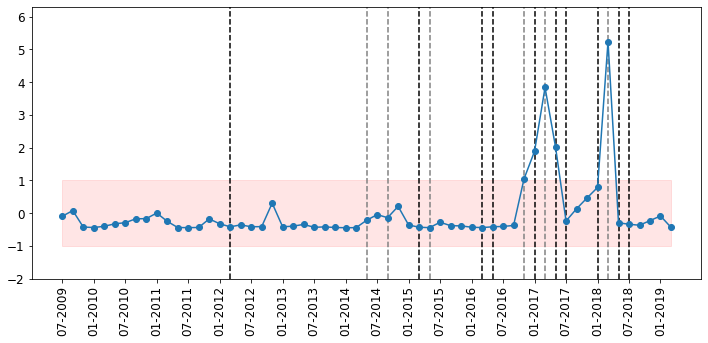}
        \caption{Proportion of posts in negative threads to posts in positive threads.}
        \label{fig:vaccine_posts_neg_pos}
    \end{subfigure}
    \begin{subfigure}{\textwidth}
    \centering
    \includegraphics[width=0.9\textwidth]{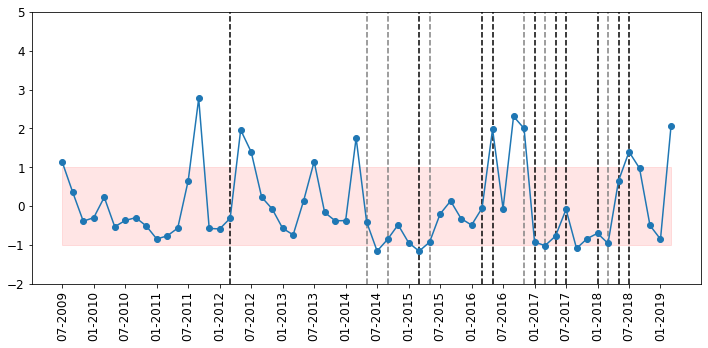}
        \caption{Proportion of posts in neutral threads to posts in positive and negative threads.}
        \label{fig:vaccine_posts_neutral}
    \end{subfigure}
    \caption{Vaccination subforum - posts in threads of each sentiment.}
    \label{fig:vaccine_posts_sentiment}
\end{figure}

\begin{figure}
    \centering
    \begin{subfigure}{\textwidth}
    \centering
    \includegraphics[width=0.93\textwidth]{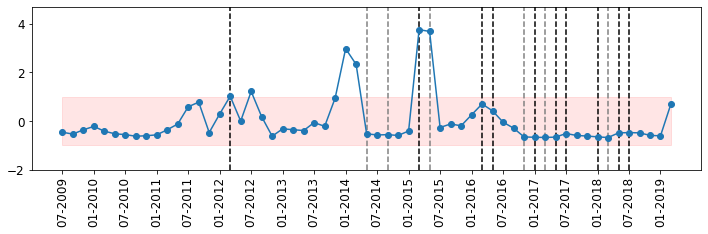}
        \caption{Proportion of users in positive threads to users in negative threads.}
        \label{fig:vaccine_users_pos_neg}
    \end{subfigure}
    \begin{subfigure}{\textwidth}
    \centering
    \includegraphics[width=0.93\textwidth]{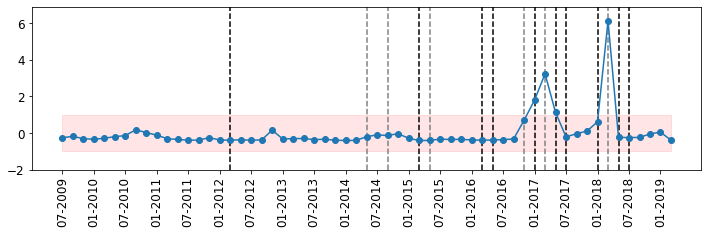}
        \caption{Proportion of users in negative threads to users in positive threads.}
        \label{fig:vaccine_users_neg_pos}
    \end{subfigure}
    \begin{subfigure}{\textwidth}
    \centering
    \includegraphics[width=0.93\textwidth]{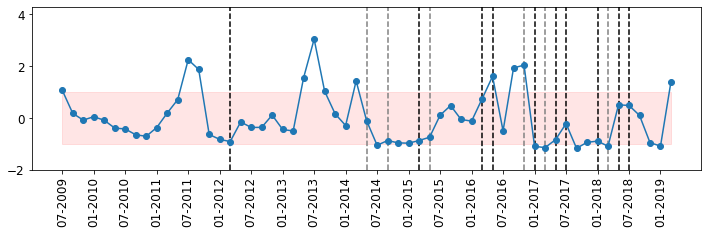}
        \caption{Proportion of users in neutral threads to users in positive and negative threads.}
        \label{fig:vaccine_users_neutral}
    \end{subfigure}
    \begin{subfigure}{\textwidth}
    \centering
    \includegraphics[width=0.93\textwidth]{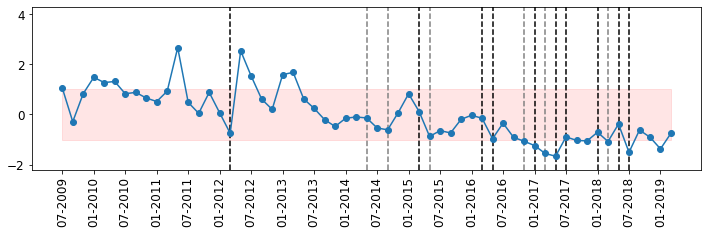}
        \caption{Proportion of users that post in multiple threads with different sentiments to users only in neutral, positive or negative threads.}
        \label{fig:vaccine_users_mult_sent}
    \end{subfigure}
    \caption{Vaccination subforum - users in threads of each sentiment.}
    \label{fig:vaccine_users_sentiment}
\end{figure}

\begin{figure}
    \centering
    \begin{subfigure}{\textwidth}
    \centering
    \includegraphics[width=0.9\textwidth]{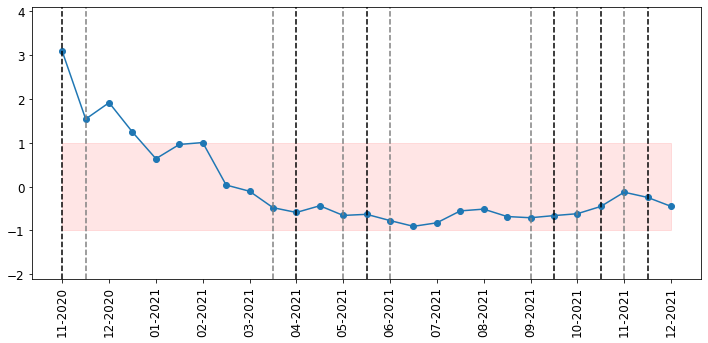}
        \caption{Proportion of positive to negative threads.}
        \label{fig:covid_thread_pos_neg}
    \end{subfigure}
    \begin{subfigure}{\textwidth}
    \centering
    \includegraphics[width=0.9\textwidth]{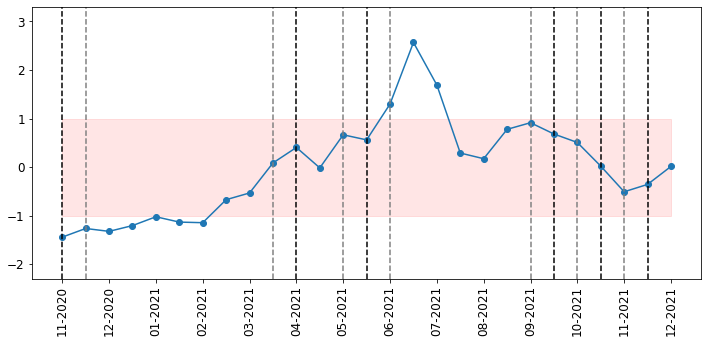}
        \caption{Proportion of negative to positive threads.}
        \label{fig:covid_thread_neg_pos}
    \end{subfigure}
    \begin{subfigure}{\textwidth}
    \centering
    \includegraphics[width=0.9\textwidth]{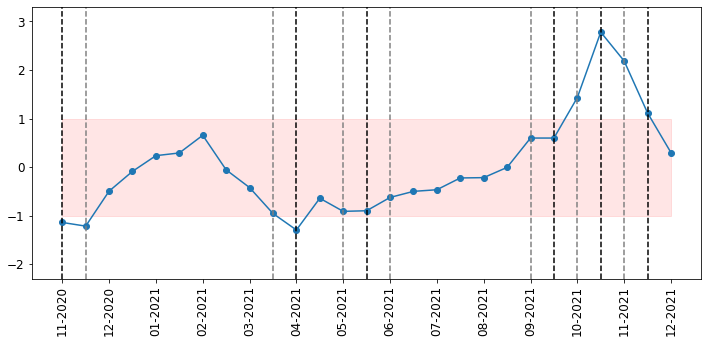}
        \caption{Proportion of neutral to positive and negative threads.}
        \label{fig:covid_thread_neutral}
    \end{subfigure}
    \caption{COVID-19 subforum - threads of each sentiment.}
    \label{fig:covid_thread_sentiment}
\end{figure}

\begin{figure}
    \centering
    \begin{subfigure}{\textwidth}
    \centering
    \includegraphics[width=0.9\textwidth]{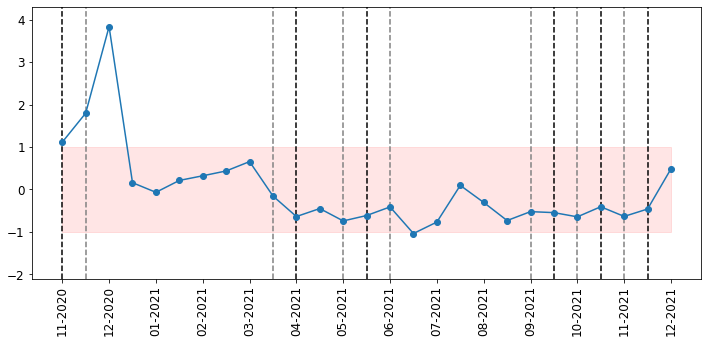}
        \caption{Proportion of posts in positive threads to posts in negative threads.}
        \label{fig:covid_posts_pos_neg}
    \end{subfigure}
    \begin{subfigure}{\textwidth}
    \centering
    \includegraphics[width=0.9\textwidth]{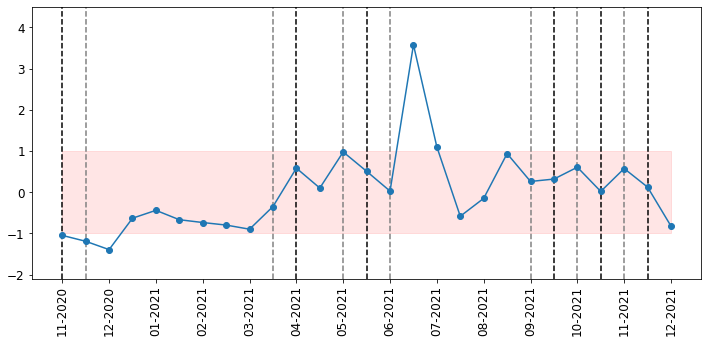}
        \caption{Proportion of posts in negative threads to posts in positive threads.}
        \label{fig:covid_posts_neg_pos}
    \end{subfigure}
    \begin{subfigure}{\textwidth}
    \centering
    \includegraphics[width=0.9\textwidth]{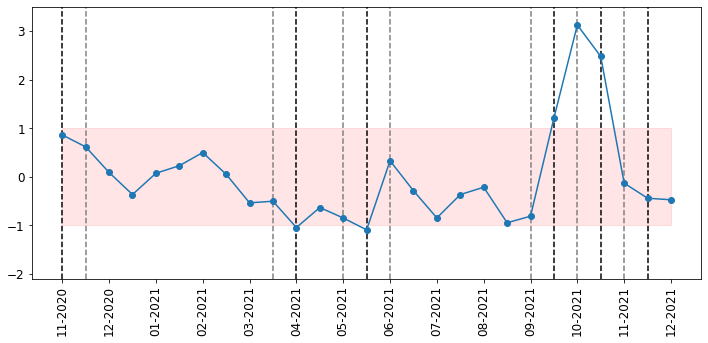}
        \caption{Proportion of posts in neutral threads to posts in positive and negative threads.}
        \label{fig:covid_posts_neutral}
    \end{subfigure}
    \caption{COVID-19 subforum - posts in threads of each sentiment.}
    \label{fig:covid_posts_sentiment}
\end{figure}

\begin{figure}
    \centering
    \begin{subfigure}{\textwidth}
    \centering
    \includegraphics[width=0.93\textwidth]{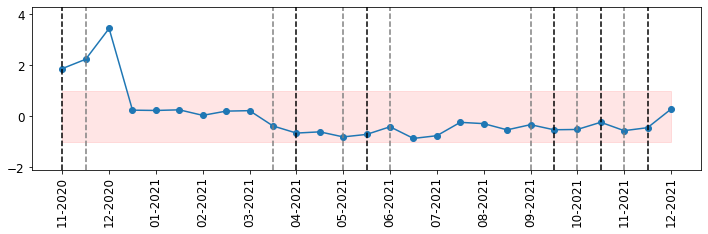}
        \caption{Proportion of users in positive threads to users in negative threads.}
        \label{fig:covid_users_pos_neg}
    \end{subfigure}
    \begin{subfigure}{\textwidth}
    \centering
    \includegraphics[width=0.93\textwidth]{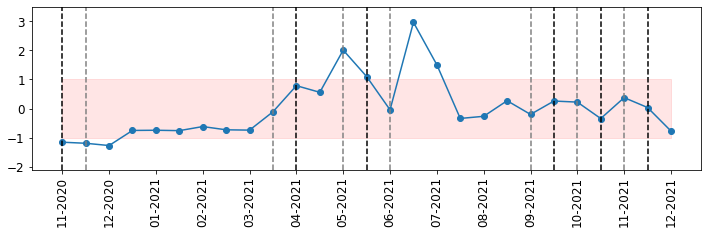}
        \caption{Proportion of users in negative threads to users in positive threads.}
        \label{fig:covid_users_neg_pos}
    \end{subfigure}
    \begin{subfigure}{\textwidth}
    \centering
    \includegraphics[width=0.93\textwidth]{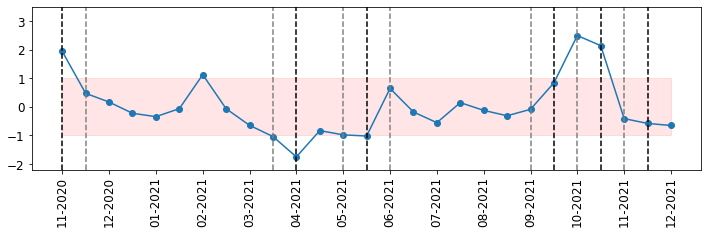}
        \caption{Proportion of users in neutral threads to users in positive and negative threads.}
        \label{fig:covid_users_neutral}
    \end{subfigure}
    \begin{subfigure}{\textwidth}
    \centering
    \includegraphics[width=0.93\textwidth]{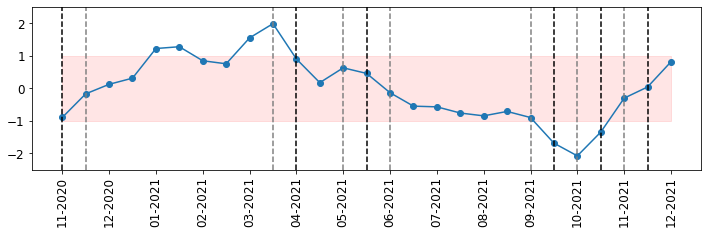}
        \caption{Proportion of users that post in multiple threads with different sentiments to users only in neutral, positive or negative threads.}
        \label{fig:covid_users_mult_sent}
    \end{subfigure}
    \caption{COVID-19 subforum - users in threads of each sentiment.}
    \label{fig:covid_users_sentiment}
\end{figure}

\begin{figure}
    \centering
    \begin{subfigure}{\textwidth}
    \centering
    \includegraphics[width=0.85\textwidth]{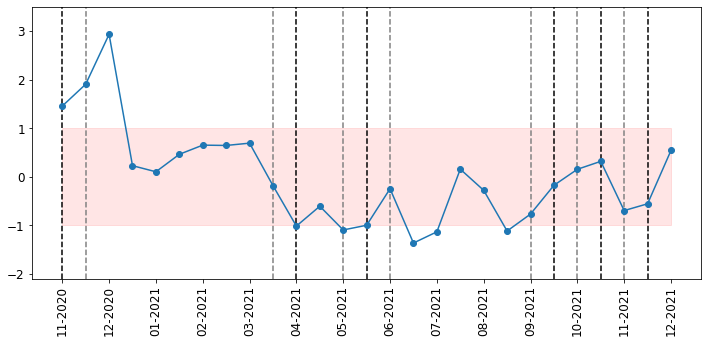}
        \caption{Proportion of positive to negative posts.}
        \label{fig:covid_posts_hand_pos_neg}
    \end{subfigure}
    \begin{subfigure}{\textwidth}
    \centering
    \includegraphics[width=0.85\textwidth]{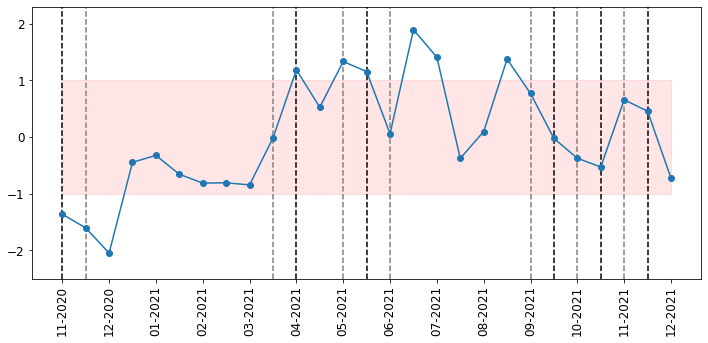}
        \caption{Proportion of negative to positive posts.}
        \label{fig:covid_posts_hand_neg_pos}
    \end{subfigure}
    \begin{subfigure}{\textwidth}
    \centering
    \includegraphics[width=0.85\textwidth]{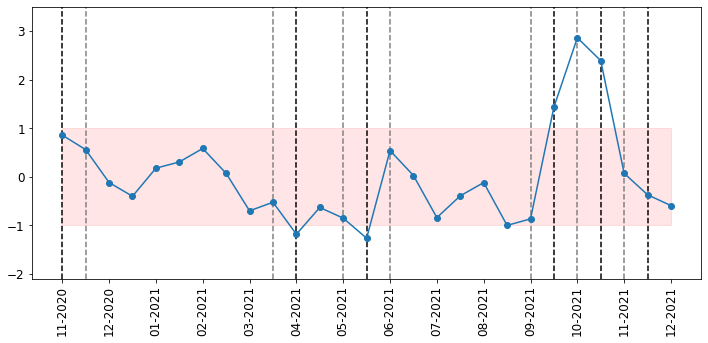}
        \caption{Proportion of neutral to positive and negative posts.}
        \label{fig:covid_posts_hand_neutral}
    \end{subfigure}
    \caption{COVID-19 subforum - posts of each sentiment, inferred from the sampled subset of data.}
    \label{fig:covid_posts_hand_sentiment}
\end{figure}

\clearpage

\end{appendices}


\bibliography{paper}

\begin{thebibliography}{73}
\providecommand{\natexlab}[1]{#1}
\providecommand{\url}[1]{{#1}}
\providecommand{\urlprefix}{URL }
\providecommand{\doi}[1]{\url{https://doi.org/#1}}
\providecommand{\eprint}[2][]{\url{#2}}
 \bibcommenthead

\bibitem[{Addawood(2018)}]{addawood2018usage}
Addawood A (2018) Usage of scientific references in mmr vaccination debates on
  twitter. In: 2018 IEEE/ACM International Conference on Advances in Social
  Networks Analysis and Mining (ASONAM), IEEE, pp 971--979

\bibitem[{Aldous et~al(1985)Aldous, Ibragimov, Jacod, and
  Aldous}]{aldous1985exchangeability}
Aldous DJ, Ibragimov IA, Jacod J, et~al (1985) Exchangeability and related
  topics. Springer

\bibitem[{Barab{\'a}si and Albert(1999)}]{barabasi1999emergence}
Barab{\'a}si AL, Albert R (1999) Emergence of scaling in random networks.
  Science 286(5439):509--512

\bibitem[{Bello-Orgaz et~al(2017)Bello-Orgaz, Hernandez-Castro, and
  Camacho}]{bello2017detecting}
Bello-Orgaz G, Hernandez-Castro J, Camacho D (2017) Detecting discussion
  communities on vaccination in twitter. Future Generation Computer Systems
  66:125--136

\bibitem[{Blankenship et~al(2018)Blankenship, Goff, Yin, Tse, Fu, Liang,
  Saroha, and Fung}]{blankenship2018sentiment}
Blankenship EB, Goff ME, Yin J, et~al (2018) Sentiment, contents, and retweets:
  A study of two vaccine-related twitter datasets. The Permanente Journal 22

\bibitem[{Blume(2006)}]{blume2006anti}
Blume S (2006) Anti-vaccination movements and their interpretations. Social
  Science \& Medicine 62(3):628--642

\bibitem[{Borgatti and Everett(1997)}]{borgatti1997network}
Borgatti SP, Everett MG (1997) Network analysis of 2-mode data. Social Networks
  19(3):243--270

\bibitem[{Breiman(2001)}]{breiman2001random}
Breiman L (2001) Random forests. Machine Learning 45(1):5--32

\bibitem[{Brown et~al(2020)Brown, Mann, Ryder, Subbiah, Kaplan, Dhariwal,
  Neelakantan, Shyam, Sastry, Askell et~al}]{brown2020language}
Brown T, Mann B, Ryder N, et~al (2020) Language models are few-shot learners.
  Advances in neural information processing systems 33:1877--1901

\bibitem[{Brunson(2013)}]{brunson2013impact}
Brunson EK (2013) The impact of social networks on parents' vaccination
  decisions. Pediatrics 131(5):e1397--e1404

\bibitem[{Buhrmester et~al(2016)Buhrmester, Kwang, and
  Gosling}]{buhrmester2016amazon}
Buhrmester M, Kwang T, Gosling SD (2016) Amazon's Mechanical Turk: A new source
  of inexpensive, yet high-quality data?, American Psychological Association,
  pp 133–--139

\bibitem[{Campbell et~al(2017)Campbell, Edwards, Letley, Bedford, Ramsay, and
  Yarwood}]{campbell2017changing}
Campbell H, Edwards A, Letley L, et~al (2017) Changing attitudes to childhood
  immunisation in english parents. Vaccine 35(22):2979--2985

\bibitem[{Collaborative et~al(2021)Collaborative, Curtis, and
  Inglesby}]{opensafely2021trends}
Collaborative TO, Curtis HJ, Inglesby Pea (2021) Trends and clinical
  characteristics of covid-19 vaccine recipients: a federated analysis of 57.9
  million patients{\textquoteright} primary care records in situ using
  opensafely. medRxiv \doi{10.1101/2021.01.25.21250356}

\bibitem[{Cortes and Vapnik(1995)}]{cortes1995support}
Cortes C, Vapnik V (1995) Support-vector networks. Machine Learning
  20(3):273--297

\bibitem[{Davies et~al(2002)Davies, Chapman, and
  Leask}]{davies2002antivaccination}
Davies P, Chapman S, Leask J (2002) Antivaccination activists on the world wide
  web. Archives of Disease in Childhood 87(1):22--25

\bibitem[{Denford et~al(2022)Denford, Mowbray, Towler, Wehling, Lasseter,
  Aml{\^o}t, Oliver, Yardley, and Hickman}]{denford2022exploration}
Denford S, Mowbray F, Towler L, et~al (2022) Exploration of attitudes regarding
  uptake of covid-19 vaccines among vaccine hesitant adults in the uk: a
  qualitative analysis. BMC Infectious Diseases 22(1):1--14

\bibitem[{DeRoo et~al(2020)DeRoo, Pudalov, and Fu}]{deroo2020planning}
DeRoo SS, Pudalov NJ, Fu LY (2020) Planning for a covid-19 vaccination program.
  JAMA 323(24):2458--2459

\bibitem[{Elliott et~al(2022)Elliott, Bodinier, Eales, Wang, Haw, Elliott,
  Whitaker, Jonnerby, Tang, Walters et~al}]{elliott2022rapid}
Elliott P, Bodinier B, Eales O, et~al (2022) Rapid increase in omicron
  infections in england during december 2021: React-1 study. Science
  375(6587):1406--1411

\bibitem[{{European Medical Agency}(2021)}]{bloodclot}
{European Medical Agency} (2021)
  \url{https://www.ema.europa.eu/en/news/astrazenecas-covid-19-vaccine-ema-finds-possible-link-very-rare-cases-unusual-blood-clots-low-blood}

\bibitem[{Ford and Alwan(2018)}]{ford2018use}
Ford AJ, Alwan NA (2018) Use of social networking sites and women's decision to
  receive vaccinations during pregnancy: A cross-sectional study in the uk.
  Vaccine 36(35):5294--5303

\bibitem[{Garcia-Beltran et~al(2022)Garcia-Beltran, {St. Denis}, and
  et~al.}]{garciabeltran2022mrna}
Garcia-Beltran WF, {St. Denis} KJ, et~al. AH (2022) mrna-based covid-19 vaccine
  boosters induce neutralizing immunity against sars-cov-2 omicron variant.
  Cell 185(3):457--466.e4.
  \urlprefix\url{https://www.sciencedirect.com/science/article/pii/S0092867421014963}

\bibitem[{{GOV.UK}(2020)}]{pfizerapproval}
{GOV.UK} (2020)
  \url{https://www.gov.uk/government/publications/regulatory-approval-of-pfizer-biontech-vaccine-for-covid-19}

\bibitem[{{GOV.UK}(2022)}]{govukvaccs}
{GOV.UK} (2022) Vaccinations in the uk.
  https://coronavirus.data.gov.uk/details/vaccinations, accessed 03-05-2022

\bibitem[{Guilbeault et~al(2018)Guilbeault, Becker, and
  Centola}]{Guilbeault:2018}
Guilbeault D, Becker J, Centola D (2018) Complex contagions: a decade in
  review, Springer International Publishing, Cham, Switzerland, pp 3--25

\bibitem[{Hoffman et~al(2019)Hoffman, Felter, Chu, Shensa, Hermann, Wolynn,
  Williams, and Primack}]{hoffman2019its}
Hoffman BL, Felter EM, Chu KH, et~al (2019) It's not all about autism: The
  emerging landscape of anti-vaccination sentiment on facebook. Vaccine
  37(16):2216--2223.
  \urlprefix\url{https://www.sciencedirect.com/science/article/pii/S0264410X19303032}

\bibitem[{Hooker(2014)}]{hooker2014measles}
Hooker BS (2014) Measles-mumps-rubella vaccination timing and autism among
  young african american boys: a reanalysis of cdc data. Translational
  Neurodegeneration 3(1):1--6

\bibitem[{House et~al(2009)House, Davies, Danon, and Keeling}]{House:2009}
House T, Davies G, Danon L, et~al (2009) A motif-based approach to network
  epidemics. Bulletin of Mathematical Biology 71(7):1693—1706

\bibitem[{Hutto and Gilbert(2014)}]{hutto2014vader}
Hutto C, Gilbert E (2014) Vader: A parsimonious rule-based model for sentiment
  analysis of social media text. In: Proceedings of the International AAAI
  Conference on Web and Social Media, pp 216--225

\bibitem[{Iacobucci(2021)}]{iacobucci2021covid}
Iacobucci G (2021) Covid-19: How is the uk’s vaccine booster programme
  faring?

\bibitem[{Jang et~al(2019)Jang, Mckeever, Mckeever, and Kim}]{jang2019social}
Jang SM, Mckeever BW, Mckeever R, et~al (2019) From social media to mainstream
  news: the information flow of the vaccine-autism controversy in the us,
  canada, and the uk. Health Communication 34(1):110--117

\bibitem[{Joulin et~al(2017)Joulin, Grave, Bojanowski, and
  Mikolov}]{joulin2017bag}
Joulin A, Grave E, Bojanowski P, et~al (2017) Bag of tricks for efficient text
  classification. In: Proceedings of the 15th Conference of the European
  Chapter of the Association for Computational Linguistics: Volume 2, Short
  Papers. Association for Computational Linguistics, pp 427--431

\bibitem[{Love et~al(2013)Love, Himelboim, Holton, and
  Stewart}]{love2013twitter}
Love B, Himelboim I, Holton A, et~al (2013) Twitter as a source of vaccination
  information: content drivers and what they are saying. American Journal of
  Infection Control 41(6):568--570

\bibitem[{Milo et~al(2002)Milo, Shen-Orr, Itzkovitz, Kashtan, Chklovskii, and
  Alon}]{milo2002network}
Milo R, Shen-Orr S, Itzkovitz S, et~al (2002) Network motifs: simple building
  blocks of complex networks. Science 298(5594):824--827

\bibitem[{Milo et~al(2004)Milo, Itzkovitz, Kashtan, Levitt, Shen-Orr,
  Ayzenshtat, Sheffer, and Alon}]{milo2004superfamilies}
Milo R, Itzkovitz S, Kashtan N, et~al (2004) Superfamilies of evolved and
  designed networks. Science 303(5663):1538--1542

\bibitem[{Mouzo and {El Pa\'{i}s}(2015)}]{diphtheria}
Mouzo J, {El Pa\'{i}s} (2015)
  \url{https://english.elpais.com/elpais/2015/06/05/inenglish/1433512717_575817.html}

\bibitem[{M{\"u}ller and Salath{\'e}(2019)}]{muller2019crowdbreaks}
M{\"u}ller M, Salath{\'e} M (2019) Crowdbreaks: tracking health trends using
  public social media data and crowdsourcing. Frontiers in Public Health 7:81

\bibitem[{M{\"u}ller and Salath{\'e}(2020)}]{muller2020addressing}
M{\"u}ller M, Salath{\'e} M (2020) Addressing machine learning concept drift
  reveals declining vaccine sentiment during the covid-19 pandemic. arXiv
  preprint arXiv:201202197

\bibitem[{Musk(2022)}]{musk2022}
Musk E (2022) {World Cup traffic hit almost 20,000 tweets per second today!
  Great work by Twitter team managing record usage.}
  \url{https://twitter.com/elonmusk/status/1595505413113323520}

\bibitem[{Newman(2001)}]{newman2001scientific}
Newman ME (2001) Scientific collaboration networks. ii. shortest paths,
  weighted networks, and centrality. Physical Review E 64(1):016,132

\bibitem[{Omer et~al(2009)Omer, Salmon, Orenstein, deHart, and
  Halsey}]{omer2009vaccine}
Omer SB, Salmon DA, Orenstein WA, et~al (2009) Vaccine refusal, mandatory
  immunization, and the risks of vaccine-preventable diseases. New England
  Journal of Medicine 360(19):1981--1988

\bibitem[{Peretti-Watel et~al(2020)Peretti-Watel, Seror, Cortaredona, Launay,
  Raude, Verger, Fressard, Beck, Legleye, l'Haridon et~al}]{peretti2020future}
Peretti-Watel P, Seror V, Cortaredona S, et~al (2020) A future vaccination
  campaign against covid-19 at risk of vaccine hesitancy and politicisation.
  The Lancet Infectious Diseases pp 769--770

\bibitem[{Piedrahita-Vald{\'e}s et~al(2021)Piedrahita-Vald{\'e}s,
  Piedrahita-Castillo, Bermejo-Higuera, Guillem-Saiz, Bermejo-Higuera,
  Guillem-Saiz, Sicilia-Montalvo, and
  Mach{\'\i}o-Regidor}]{piedrahita2021vaccine}
Piedrahita-Vald{\'e}s H, Piedrahita-Castillo D, Bermejo-Higuera J, et~al (2021)
  Vaccine hesitancy on social media: Sentiment analysis from june 2011 to april
  2019. Vaccines 9(1):28

\bibitem[{Pouwels et~al(2021)Pouwels, Pritchard, Matthews, Stoesser, Eyre,
  Vihta, House, Hay, Bell, Newton et~al}]{pouwels2021effect}
Pouwels KB, Pritchard E, Matthews PC, et~al (2021) Effect of delta variant on
  viral burden and vaccine effectiveness against new sars-cov-2 infections in
  the uk. Nature Medicine 27(12):2127--2135

\bibitem[{Pr{\v{z}}ulj(2007)}]{prvzulj2007biological}
Pr{\v{z}}ulj N (2007) Biological network comparison using graphlet degree
  distribution. Bioinformatics 23(2):e177--e183

\bibitem[{Radzikowski et~al(2016)Radzikowski, Stefanidis, Jacobsen, Croitoru,
  Crooks, and Delamater}]{radzikowski2016measles}
Radzikowski J, Stefanidis A, Jacobsen KH, et~al (2016) The measles vaccination
  narrative in twitter: a quantitative analysis. JMIR Public Health and
  Surveillance 2(1):e5059

\bibitem[{Razai et~al(2021)Razai, Oakeshott, Esmail, Wiysonge, Viswanath, and
  Mills}]{razai2021covid19}
Razai MS, Oakeshott P, Esmail A, et~al (2021) Covid-19 vaccine hesitancy: the
  five cs to tackle behavioural and sociodemographic factors. Journal of the
  Royal Society of Medicine 114(6):295--298. \doi{10.1177/01410768211018951},
  \urlprefix\url{https://doi.org/10.1177/01410768211018951}, pMID: 34077688,
  {\href{https://arxiv.org/abs/https://doi.org/10.1177/01410768211018951}{{https://arxiv.org/abs/https://doi.org/10.1177/01410768211018951}}}

\bibitem[{Ribeiro et~al(2021)Ribeiro, Paredes, Silva, Aparicio, and
  Silva}]{ribeiro2019survey}
Ribeiro P, Paredes P, Silva MEP, et~al (2021) A survey on subgraph counting:
  concepts, algorithms, and applications to network motifs and graphlets. ACM
  Computing Surveys 54(2). \doi{10.1145/3433652},
  \urlprefix\url{https://doi.org/10.1145/3433652}

\bibitem[{Rier(2007)}]{rier2007impact}
Rier DA (2007) The impact of moral suasion on internet hiv/aids support groups:
  evidence from a discussion of seropositivity disclosure ethics. Health
  Sociology Review 16(3-4):237--247

\bibitem[{Ritchie et~al(2014)Ritchie, Berthouze, House, and
  Kiss}]{Ritchie:2014}
Ritchie M, Berthouze L, House T, et~al (2014) Higher-order structure and
  epidemic dynamics in clustered networks. Journal of Theoretical Biology
  348:21--32

\bibitem[{Ritchie et~al(2016)Ritchie, Berthouze, and Kiss}]{Ritchie:2016}
Ritchie M, Berthouze L, Kiss IZ (2016) Generation and analysis of networks with
  a prescribed degree sequence and subgraph family: higher-order structure
  matters. Journal of Complex Networks 5(1):1--31

\bibitem[{Roberts(2022)}]{mumsnet}
Roberts J (2022) Mumsnet. \url{www.mumsnet.com}, accessed: 01-05-2022

\bibitem[{Robertson et~al(2021)Robertson, Reeve, Niedzwiedz, Moore, Blake,
  Green, Katikireddi, and Benzeval}]{robertson2021predictors}
Robertson E, Reeve KS, Niedzwiedz CL, et~al (2021) Predictors of covid-19
  vaccine hesitancy in the uk household longitudinal study. Brain, Behavior,
  and Immunity 94:41--50

\bibitem[{Romijn et~al(2015)Romijn, Nuall{\'a}in, and
  Torenvliet}]{romijn2015discovering}
Romijn L, Nuall{\'a}in B{\'O}, Torenvliet L (2015) Discovering motifs in
  real-world social networks. In: International Conference on Current Trends in
  Theory and Practice of Informatics, Springer, pp 463--474

\bibitem[{Rubner et~al(1998)Rubner, Tomasi, and Guibas}]{rubner1998metric}
Rubner Y, Tomasi C, Guibas LJ (1998) A metric for distributions with
  applications to image databases. In: Sixth International Conference on
  Computer Vision, IEEE, pp 59--66

\bibitem[{Salath{\'e} and Khandelwal(2011)}]{salathe2011assessing}
Salath{\'e} M, Khandelwal S (2011) Assessing vaccination sentiments with online
  social media: implications for infectious disease dynamics and control. PLoS
  Computational Biology 7(10):e1002,199

\bibitem[{Salath{\'e} et~al(2013)Salath{\'e}, Vu, Khandelwal, and
  Hunter}]{salathe2013dynamics}
Salath{\'e} M, Vu DQ, Khandelwal S, et~al (2013) The dynamics of health
  behavior sentiments on a large online social network. EPJ Data Science
  2(1):1--12

\bibitem[{Scannell et~al(2021)Scannell, Desens, Guadagno, Tra, Acker, Sheridan,
  Rosner, Mathieu, and Fulk}]{scannell2021covid}
Scannell D, Desens L, Guadagno M, et~al (2021) Covid-19 vaccine discourse on
  twitter: A content analysis of persuasion techniques, sentiment and
  mis/disinformation. Journal of Health Communication 26(7):443--459

\bibitem[{Schmidt et~al(2018)Schmidt, Zollo, Scala, Betsch, and
  Quattrociocchi}]{schmidt2018social}
Schmidt AL, Zollo F, Scala A, et~al (2018) Polarization of the vaccination
  debate on facebook. Vaccine 36(25):3606--3612.
  \urlprefix\url{https://www.sciencedirect.com/science/article/pii/S0264410X18306601}

\bibitem[{Silva et~al(2023)Silva, Gaunt, Ospina-Forero, Jay, and
  House}]{silva2022comparing}
Silva ME, Gaunt RE, Ospina-Forero L, et~al (2023) Comparing directed networks
  via denoising graphlet distributions. Journal of Complex Networks
  11(2):cnad006

\bibitem[{Singanayagam et~al(2022)Singanayagam, Hakki, and
  et~al.}]{singanayagam2022community}
Singanayagam A, Hakki S, et~al. JD (2022) Community transmission and viral load
  kinetics of the sars-cov-2 delta (b.1.617.2) variant in vaccinated and
  unvaccinated individuals in the uk: a prospective, longitudinal, cohort
  study. The Lancet Infectious Diseases 22(2):183--195.
  \doi{https://doi.org/10.1016/S1473-3099(21)00648-4}

\bibitem[{Skea et~al(2008)Skea, Entwistle, Watt, and
  Russell}]{skea2008avoiding}
Skea ZC, Entwistle VA, Watt I, et~al (2008) `avoiding harm to others'
  considerations in relation to parental measles, mumps and rubella (mmr)
  vaccination discussions -- an analysis of an online chat forum. Social
  Science \& Medicine 67(9):1382--1390.
  \doi{https://doi.org/10.1016/j.socscimed.2008.07.006}

\bibitem[{Skeppstedt et~al(2017)Skeppstedt, Kerren, and
  Stede}]{skeppstedt2017automatic}
Skeppstedt M, Kerren A, Stede M (2017) Automatic detection of stance towards
  vaccination in online discussion forums. In: Proceedings of the International
  Workshop on Digital Disease Detection using Social Media 2017 (DDDSM-2017),
  pp 1--8

\bibitem[{Skeppstedt et~al(2018)Skeppstedt, Kerren, and
  Stede}]{skeppstedt2018vaccine}
Skeppstedt M, Kerren A, Stede M (2018) Vaccine hesitancy in discussion forums:
  computer-assisted argument mining with topic models. In: MIE, pp 366--370

\bibitem[{Torjesen(2021)}]{torjesen2021covid}
Torjesen I (2021) Covid-19: Delta variant is now uk’s most dominant strain
  and spreading through schools

\bibitem[{Touvron et~al(2023)Touvron, Lavril, Izacard, Martinet, Lachaux,
  Lacroix, Rozi{\`e}re, Goyal, Hambro, Azhar et~al}]{touvron2023llama}
Touvron H, Lavril T, Izacard G, et~al (2023) Llama: Open and efficient
  foundation language models. arXiv preprint arXiv:230213971

\bibitem[{Tregoning et~al(2021)Tregoning, Flight, Higham, Wang, and
  Pierce}]{tregoning2021progress}
Tregoning JS, Flight KE, Higham SL, et~al (2021) Progress of the covid-19
  vaccine effort: viruses, vaccines and variants versus efficacy, effectiveness
  and escape. Nature Reviews Immunology 21(10):626--636

\bibitem[{Wegner et~al(2018)Wegner, Ospina-Forero, Gaunt, Deane, and
  Reinert}]{wegner2017identifying}
Wegner AE, Ospina-Forero L, Gaunt RE, et~al (2018) Identifying networks with
  common organizational principles. Journal of Complex Networks 6(6):887--913

\bibitem[{Weller et~al(2016)Weller, White et~al}]{weller2016content}
Weller R, White S, et~al (2016) A content analysis of online forum discussion
  about measles, mumps and rubella (mmr) vaccination between 2004 and 2015.
  International Journal of Pharmacy Practice 24:25--25

\bibitem[{Widmer and Kubat(1996)}]{widmer1996learning}
Widmer G, Kubat M (1996) Learning in the presence of concept drift and hidden
  contexts. Machine Learning 23(1):69--101

\bibitem[{Wise(2021)}]{wisen2021covid19}
Wise J (2021) Covid-19: rare immune response may cause clots after astrazeneca
  vaccine, say researchers. BMJ 373. \doi{10.1136/bmj.n954},
  \urlprefix\url{https://www.bmj.com/content/373/bmj.n954},
  {\href{https://arxiv.org/abs/https://www.bmj.com/content/373/bmj.n954.full.pdf}{{https://arxiv.org/abs/https://www.bmj.com/content/373/bmj.n954.full.pdf}}}

\bibitem[{{World Health Organisation}(2019)}]{who2019}
{World Health Organisation} (2019) Ten threats to global health in 2019.
  Https://www.who.int/news-room/spotlight/ten-threats-to-global-health-in-2019

\bibitem[{Yuan et~al(2019)Yuan, Schuchard, and Crooks}]{yuan2019examining}
Yuan X, Schuchard RJ, Crooks AT (2019) Examining emergent communities and
  social bots within the polarized online vaccination debate in twitter. Social
  media+ society 5(3):2056305119865,465

\bibitem[{Zhongbao and Changshui(2003)}]{zhongbao2003reply}
Zhongbao K, Changshui Z (2003) Reply networks on a bulletin board system.
  Physical Review E 67(3):036,117

\end{thebibliography}


\end{document}